\documentclass[a4paper,12pt]{article}
\usepackage{a4,amsmath,amssymb,axodraw,slashed,epsf}
\usepackage[square,comma,numbers,sort&compress]{natbib}
\usepackage[dvips]{color}
\usepackage{hyperref}
\usepackage{axodraw}
\usepackage{ulem}

\begin{document}


\newcommand{\bed}{\begin{displaymath}}
\newcommand{\eed}{\end{displaymath}}
\newcommand{\beq}{\begin{equation}}
\newcommand{\eeq}{\end{equation}}
\newcommand{\bea}{\begin{eqnarray}}
\newcommand{\eea}{\end{eqnarray}}
\newcommand{\tgb}{{\rm tg}\beta}
\newcommand{\tga}{{\rm tg}\alpha}
\newcommand{\stgb}{{\rm tg}^2\beta}
\newcommand{\sia}{\sin\alpha}
\newcommand{\coa}{\cos\alpha}
\newcommand{\sib}{\sin\beta}
\newcommand{\cob}{\cos\beta}
\newcommand{\MS}{\overline{\rm MS}}
\newcommand{\st}{\tilde{t}}
\newcommand{\sgl}{\tilde{g}}
\newcommand{\Dmb}{\ensuremath{\Delta_b}}

\newcommand{\tb}{{\rm tg}\beta}
\newcommand{\gluino}{\widetilde{g}}
\newcommand{\squark}{\tilde{q}}
\newcommand{\Mq}[1]{m_{\tilde{q}_{#1}}}
\newcommand{\mq}{m_{q}}
\newcommand{\Aq}{A_q}
\newcommand{\mix}{\widetilde{\theta}}
\newcommand{\Lag}{\mathcal{L}}
\newcommand{\ls}{\lambda_{s}}
\newcommand{\lb}{\lambda_{b}}
\newcommand{\lt}{\lambda_t}
\newcommand{\Deltamb}{\Delta_b}
\newcommand{\DeltambQCD}{\Delta_b^{QCD}}
\newcommand{\DeltambELW}{\Delta_b^{elw}}
\newcommand{\mb}{m_{b}}
\newcommand{\mt}{m_{t}}
\newcommand{\bR}{b_R}
\newcommand{\bL}{b_L}
\newcommand{\sR}{s_R}
\newcommand{\sL}{s_L}
\newcommand{\At}{A_t}
\newcommand{\Sb}{\Sigma_b}
\newcommand{\Ao}[1]{A_0(#1)}
\newcommand{\Co}[5]{C_0(#1,#2;#3,#4,#5)}
\newcommand{\as}{\alpha_s}
\newcommand{\Mg}{m_{\tilde{g}}}
\newcommand{\Mb}[1]{m_{\tilde{b}_{#1}}}
\newcommand{\Mt}[1]{m_{\tilde{t}_{#1}}}
\newcommand{\NL}{\nonumber\\}
\newcommand{\bs}{\widetilde{b}}
\newcommand{\str}{\widetilde{s}}
\newcommand{\bsL}{\bs_L}
\newcommand{\bsR}{\bs_R}
\newcommand{\strL}{\str_L}
\newcommand{\strR}{\str_R}
\newcommand{\ts}{\widetilde{t}}
\newcommand{\tsR}{\widetilde{t}_R}
\newcommand{\tsL}{\widetilde{t}_L}
\newcommand{\higgsino}{\widetilde{h}}
\newcommand{\stopx}{\tilde{t}}
\newcommand{\sbottom}{\tilde{b}}
\newcommand{\strange}{\tilde{s}}
\newcommand{\gh}{g_b^h}
\newcommand{\gH}{g_b^H}
\newcommand{\gA}{g_b^A}
\newcommand{\ght}{\tilde{g}_b^h}
\newcommand{\gHt}{\tilde{g}_b^H}
\newcommand{\gAt}{\tilde{g}_b^A}
\newcommand{\stb}{{\rm tg}^2\beta}
\newcommand{\Order}[1]{{\cal{O}}\left(#1\right)}
\newcommand{\Msusy}{M_{SUSY}}
\newcommand{\gluon}{g}
\newcommand{\sttop}{\widetilde{t}}
\newcommand{\CF}{C_F}
\newcommand{\CA}{C_A}
\newcommand{\TR}{T_R}
\newcommand{\Tr}[1]{{\rm Tr}\left[#1\right]}
\newcommand{\dk}{\frac{d^nk}{(2\pi)^n}}
\newcommand{\dqq}{\frac{d^nq}{(2\pi)^n}}
\newcommand{\partderiv}[1]{\frac{\partial}{\partial#1}}
\newcommand{\kmu}{k_{\mu}}
\newcommand{\qmu}{q_{\mu}}
\newcommand{\eUV}{\epsilon}
\newcommand{\dMq}[1]{\delta\Mq{#1}}
\newcommand{\gs}{g_s}
\newcommand{\MSbar}{\overline{\rm MS}}
\newcommand{\gPhit}{\tilde{g}_b^{\Phi}}
\newcommand{\muR}{\mu_{R}}
\newcommand{\bbbar}{b\bar{b}}
\newcommand{\MPhi}{M_{\Phi}}
\newcommand{\dlt}{\delta\lambda_t}
\newcommand{\GF}{{\rm G_F}}
\newcommand{\gPhi}{g_b^{\Phi}}
\newcommand{\mbMS}{\overline{m}_{b}}
\newcommand{\asrun}[1]{\as(#1)}
\newcommand{\NF}{N_F}
\newcommand{\gtPhi}{g_t^{\Phi}}
\newcommand{\Log}[1]{\log\left(#1\right)}
\newcommand{\smallaeff}{small $\alpha_{eff}$}
\newcommand{\ra}{\rightarrow}
\newcommand{\tautau}{\tau^+\tau^-}

\newcommand{\Quark}[4]{\ArrowLine(#1,#2)(#3,#4)}
\newcommand{\Gluino}[5]{\Gluon(#1,#2)(#3,#4){3}{#5}\Line(#1,#2)(#3,#4)}
\newcommand{\CrossedCircle}[6]{\BCirc(#1,#2){5}\Line(#3,#6)(#5,#4)\Line(#5,#6)(#3,#4)}
\newcommand{\Higgsino}[5]{\Photon(#1,#2)(#3,#4){3}{#5}\Line(#1,#2)(#3,#4)}
\newcommand{\myGluon}[5]{\Gluon(#1,#2)(#3,#4){3}{#5}}
\newcommand{\Squark}[4]{\DashArrowLine(#1,#2)(#3,#4){2}}
\newcommand{\Squarknoarrow}[4]{\DashLine(#1,#2)(#3,#4){2}}
\newcommand{\Quarknoarrow}[4]{\ArrowLine(#1,#2)(#3,#4)}

\newcommand{\lsim}{\raisebox{-0.13cm}{~\shortstack{$<$ \\[-0.07cm] $\sim$}}~}
\newcommand{\gsim}{\raisebox{-0.13cm}{~\shortstack{$>$ \\[-0.07cm] $\sim$}}~}


\vspace*{-2.5cm}
\begin{flushright}
{\small LHCHWG-2025-014} \\[0.5cm]
\end{flushright}
\begin{center}
{\large \sc Charged Higgs Pairs at the LHC: A NLO Analysis}
\end{center}

\begin{center}

{\sc M.~Ahmed$^{1}$, L.~Biermann$^1$, H.~Ita$^{1,2}$,
I.~Rusetski$^{1,3}$, M.~Spira$^1$ and Y.~Yedelkina$^1$} \\[0.8cm]

\begin{small}
{\it \small
$^1$ PSI Center for Neutron and Muon Sciences, CH--5232 Villigen PSI,
Switzerland \\
$^2$ Department of Astrophysics, University of Zurich,
Winterthurerstrasse 190, 8057 Zurich, Switzerland \\
$^3$ Albert Einstein Center for Fundamental Physics, Institute for
Theoretical Physics, University of Bern, Sidlerstrasse 5, 3012 Bern,
Switzerland}
\end{small}
\end{center}


\begin{abstract}
\noindent
Charged Higgs-boson pair production at hadron colliders yields
information about the trilinear couplings involving charged Higgs fields in extensions of the Standard Model (SM). We consider the type-I
two-Higgs doublet model (2HDM) extension and obtain next-to-leading order QCD predictions for the charged-Higgs pair production ($H^+H^-$ production).
All production modes, i.e. Drell--Yan-like production, gluon fusion and vector-boson fusion are included in the analysis. We determine uncertainties
originating from the scale dependence, the parton-density functions and
strong coupling $\alpha_s$ at the LHC. 
We observe that the QCD corrections lead to a significant
reduction of the relative scale dependences. These improved predictions will allow 
for a quantitative interpretation of
experimental measurements, in case that charged Higgs states will be observed.

\end{abstract}


\section{Introduction}
With the discovery of a scalar resonance at the LHC \cite{discovery}
which is compatible with the Standard Model (SM) Higgs boson
\cite{higgs,couplings} the theory of weak interactions is complete and
renormalizable, if this scalar
resonance is indeed the SM Higgs boson with SM couplings. However, since several couplings of the
scalar resonance to the known matter fields and itself are not yet or not yet sufficiently well probed
experimentally, there is ample room for deviations from the SM. In addition,
However, since the SM does not explain
several observations such as Dark Matter and Dark Energy as well as the
matter-antimatter asymmetry of the Universe, it is considered
incomplete and extensions beyond the SM (BSM) are expected. 
The search for BSM physics is concerned with two directions,
{\it (i)} the search for deviations of the properties of SM
particles from the SM prediction and {\it (ii)} the search for new particles.
Since most of the known particle properties have been tested down to the
per-cent or per-mille level, 
it is natural to focus BSM analyses on the properties of the discovered scalar resonance with a mass of about 125~GeV. Its properties are confirmed to agree with the SM Higgs boson at the 5--10\% level which leaves ample room for potential deviations from the SM.
Alternatively, searches for SM extensions including extended Higgs, gauge-boson and fermion
sectors are actively pursued at the high-energy collider experiments at the LHC.
In this work, we concentrate on extensions of the Higgs sector and we will
focus entirely on the production modes of charged Higgs states that
arise naturally in extended Higgs sectors. As a particular example,
we focus on the Two-Higgs Doublet Model (2HDM) of type I~\cite{Lee:1973iz,Branco:2011iw}.

Charged Higgs bosons are dominately produced via the associated production with top and bottom quarks, $q \bar{q},\,gg \to H^- \,\bar{b} \, t$ and the charged-conjugate process. This process is known up to NLO-QCD in the four- and five-flavour schemes~\cite{ch4fs,ch5fs,proch}. The other single charged-Higgs production processes are subleading, c.f.~\cite{review} and references therein.

This work addresses charged Higgs-pair production and adds to the presently available results for Drell--Yan-like production $q\bar{q}\rightarrow H^+H^-$ at next-to-leading order (NLO) \cite{plehn}\footnote{Recently, also a calculation of the Drell--Yan-like process for charged-Higgs pair production at \textit{approximate} N$^2$LO and N$^3$LO was published \cite{Kidonakis:2024gcj}.} and for the gluon-fusion process $gg\to H^+H^-$ at NLO QCD \cite{Hespel:2014sla} by a new and complete derivation in the heavy-top limit (HTL) and by extending the vector-boson
fusion (VBF) $qq\to qq H^+H^-$ \cite{moretti} to the NLO level as well. We perform a
detailed analysis of all related uncertainties. Section \ref{sc:2hdm}
introduces the 2HDM of type I and the set-up and scenarios used in our
work. In Section \ref{sc:dy} we discuss the Drell--Yan-like process at
NLO, while in Section \ref{sc:ggf} we elaborate on the loop-induced
gluon-fusion process $gg\to H^+H^-$ and discuss the NLO QCD corrections.
Section \ref{sc:vbf} describes our calculation of the VBF process and
the related QCD corrections at NLO. In Section \ref{sc:results}, we
present our results with the full uncertainties of the combined scale
and PDF+$\alpha_s$ errors. In Section \ref{sc:conclusions}, we conclude.

\section{The Two-Higgs Doublet Model} \label{sc:2hdm}
The 2HDM extends the SM by a second Higgs isospin-doublet with the same
hypercharge \cite{Lee:1973iz,Branco:2011iw}. We elaborate on the 2HDM
version with a softly broken $\mathbb{Z}_2$ symmetry which transforms
the two Higgs doublets $\Phi_{1,2}$ as $\Phi_1 \to -\Phi_1$ and $\Phi_2
\to \Phi_2$.  Introducing two $SU(2)_L$ Higgs doublets with hypercharge
$Y=+1$, the most general $SU(2)_L \times U(1)_Y$ invariant Higgs
potential with a softly broken $\mathbb{Z}_2$ symmetry reads
\bea
V &=& m_{11}^2 |\Phi_1|^2 + m_{22}^2 |\Phi_2|^2 - m_{12}^2 (\Phi_1^\dagger
\Phi_2 + h.c.)
 + \frac{\lambda_1}{2} (\Phi_1^\dagger \Phi_1)^2 + \frac{\lambda_2}{2}
(\Phi_2^\dagger \Phi_2)^2 + \lambda_3 (\Phi_1^\dagger \Phi_1)
(\Phi_2^\dagger \Phi_2) \nonumber \\
& + & \lambda_4 (\Phi_1^\dagger \Phi_2) (\Phi_2^\dagger \Phi_1) +
\frac{\lambda_5}{2} [(\Phi_1^\dagger \Phi_2)^2 + h.c.] \;.
\eea
We are working in the CP-conserving 2HDM, so that the three squared mass
parameters, $m_{11}^2$, $m_{22}^2$ and $m_{12}^2$, and the five coupling
parameters $\lambda_{i}$ ($i=1,...,5$) are real. A discrete $\mathbb{Z}_2$
symmetry (softly broken by the term proportional to $m_{12}^2$) is
introduced so that tree-level flavour-changing neutral currents (FCNC)
are absent. When the $\mathbb{Z}_2$ symmetry is extended to the fermion
sector, the families of same-charge fermions are enforced to couple to a
single doublet so that tree-level FCNCs will be eliminated
\cite{Branco:2011iw, Glashow:1976nt}. This allows for four different
types of fermionic doublet couplings. The associated types of the 2HDM
are named type~I, type~II, lepton-specific and flipped. The emerging
couplings of the fermions normalized to the SM couplings can be found in Ref.~\cite{Branco:2011iw}.

After electroweak symmetry breaking, the Higgs isospin-doublets $\Phi_i$
$(i=1,2)$ can be decomposed in
their vacuum expectation values (vevs) $v_i$, the charged complex fields 
$\phi_i^+$, and the real neutral CP-even ($\rho_i$) and CP-odd
($\eta_i$) fields,
\beq
\Phi_1 = \left(
\begin{array}{c}
\phi_1^+ \\[0.3cm]
\displaystyle \frac{v_1 + \rho_1 + i \eta_1}{\sqrt{2}}
\end{array}
\right) \qquad \mbox{and} \qquad
\Phi_2 = \left(
\begin{array}{c}
\phi_2^+ \\[0.3cm]
\displaystyle \frac{v_2 + \rho_2 + i \eta_2}{\sqrt{2}}
\end{array}
\right) \;.\label{eq:vevexpansion}
\eeq
The mass matrices can be derived from the bilinear Higgs-field terms in
the potential. The diagonalization is achieved by the orthogonal
transformations
\bea
\left( \begin{array}{c} \rho_1 \\ \rho_2 \end{array} \right) &=&
R(\alpha) \left( \begin{array}{c} H \\ h \end{array} \right)  \; ,
\label{eq:diagHh} \nonumber \\
\left( \begin{array}{c} \eta_1 \\ \eta_2 \end{array} \right) &=&
R(\beta) \left( \begin{array}{c} G^0 \\ A \end{array} \right)  \;
, \label{eq:diagGA} \nonumber \\
\left( \begin{array}{c} \phi_1^\pm \\ \phi^\pm_2 \end{array} \right) &=&
R(\beta) \left( \begin{array}{c} G^\pm \\ H^\pm \end{array}
\right) \label{eq:diagGHpm}
\;.
\eea
This results in five physical Higgs states, two neutral light and heavy
CP-even, $h,H$, a neutral CP-odd, $A$, and two charged Higgs bosons,
$H^\pm$, where by definition, $M_{h} < M_H$. The massless would-be
Goldstone bosons $G^\pm$ and $G^0$ are absorbed by the massive gauge
bosons $W,Z$ giving rise to their longitudinal components.  The
corresponding rotation matrices involve the mixing angles $\theta =
\alpha$ and $\beta$, respectively,
\beq
R(\theta) = \left( \begin{array}{cc} \cos \theta & - \sin
    \theta \\ \sin \theta & \cos \theta \end{array} \right) \;.
\eeq
The mixing angle $\beta$ is related to the two vevs,
\beq
\tan \beta = \frac{v_2}{v_1} \;, \label{eq:tanbetadef}
\eeq
with $v_1^2 + v_2^2 = v^2 = 1/(\sqrt{2} G_F) \approx (246 \mbox{
GeV})^2$ determined by the Fermi constant $G_F$. Introducing the
auxiliary parameters
\beq
M^2 \equiv \frac{m_{12}^2}{s_\beta c_\beta}, \qquad
\lambda_{345} \equiv \lambda_3 + \lambda_4 + \lambda_5 \;,
\eeq
the mixing angle $\alpha$ can be derived as \cite{Kanemura:2004mg}
\beq
\tan 2\alpha = \frac{s_{2\beta} (M^2- \lambda_{345} v^2)}{c_\beta^2
  (M^2-\lambda_1 v^2) -s_\beta^2 (M^2-\lambda_2 v^2)}
\;,  \label{eq:alphadef}
\eeq
adopting the short-hand notation $s_x \equiv \sin x$, $c_x \equiv \cos x$.

Using the minimum conditions of the potential, the following set of
seven independent input parameters of the model can be used, 
\beq
M_h,\; M_{H^\pm},\; \tan\beta, \; \lambda_1, \;
\lambda_3, \; \lambda_4, \; \lambda_5 \;,
\eeq
next to the SM vev derived from the Fermi constant.
In our analysis, we focus on the 2HDM type I, where the couplings of each physical Higgs boson to the up- and down-type fermions are equal. We study
two scenarios, one for small and one for large values of
$\text{tan}\beta$. The charged Higgs mass $M_{H^\pm}$ for each scenario
is varied between $300\,\text{GeV}$ and $2\,\text{TeV}$, while the
remaining five input parameters,
$\{M_h,\,\lambda_1,\,\lambda_3,\,\lambda_4,\,\lambda_5\}$, are chosen to
be constant. The variation of $M_{H^\pm}$
consequently implies a variation in the dependent parameters, i.e.~of
the mass parameters, remaining physical BSM masses and the mixing
angle~$\{m_{11}^2,\,m_{22}^2,\,m_{12}^2,\,M_H,\,M_A,\,\alpha\}$. For
each value of $M_{H^\pm}$ this new set of dependent parameters is determined via an iteration to ensure that the value of the quartic coupling $\lambda_2$ is constant too. 
By fixing the quartic couplings we ensure that the chosen
scenarios are able to fulfill all relevant theoretical and experimental
constraints over a maximal variation of the charged Higgs mass which we
check for with {\tt ScannerS}~\cite{scanners} linked with {\tt
HiggsTools~\cite{higgstools}}\footnote{The theoretical contraints which we apply by using {\tt ScannerS} are perturbative unitarity, boundedness from below and absolute stability at tree level, see~Ref.~\cite{scanners} for all details. Using {\tt ScannerS}, we furthermore require our scenarios to be consistent with experimental data, i.e.~electroweak precision data, Higgs data (via a link to {\tt HiggsTools}~\cite{higgstools}) and flavour constraints, again cf.~Ref.~\cite{scanners}.}. Our chosen scenario for
$\text{tan}\beta=2$ is constrained by electroweak precision data and
flavour observables to $M_{H^\pm}\gtrsim 344\,\text{GeV}$, while the
scenario with $\text{tan}\beta=20$ is allowed by all relevant
constraints over the full charged Higgs mass range.

The two scenarios of the 2HDM type I we use in our numerical
analysis are given by the following set of input parameters \\ \\
{\it \underline{scenario 1:}}
\begin{equation}
\begin{array}{lcllcllcl}
\;\; \tan\beta & = & 2, \;\; \qquad & \lambda_1 & = & 6.499, \qquad &
\lambda_3 & = & -1.459\,, \\
& & & \lambda_4 & = & 1.106, \qquad & \lambda_5 & = & -0.3621\,, \\
\end{array} \label{eq:scenario1} \\ \\
\end{equation} 
{\it \underline{scenario 2:}}
\begin{equation}
\begin{array}{lcllcllcl}
\tan\beta & = & 20, \qquad & \lambda_1 & = & 2.651, \qquad &
\lambda_3 & = & 0.2799\,, \\
& & & \lambda_4 & = & 2.392, \qquad & \lambda_5 & = & 2.412 \, , \\
\end{array} \label{eq:scenario2}
\end{equation} 
while the light scalar Higgs mass is fixed as $M_h = 125$ GeV.

\section{Drell--Yan like $H^+H^-$ Production} \label{sc:dy}
\begin{figure}[hbt]
\SetScale{0.7}
\begin{picture}(160,100)(-60,-20)
\ArrowLine(0,100)(50,50)
\ArrowLine(50,50)(0,0)
\Photon(50,50)(100,50){3}{5}
\DashLine(100,50)(150,100){5}
\DashLine(100,50)(150,0){5}
\put(-10,-2){$\bar q$}
\put(-10,68){$q$}
\put(42,45){$\gamma,Z$}
\put(108,67){$H^+$}
\put(108,-4){$H^-$}
\put(52,-25){$(a)$}
\end{picture}
\begin{picture}(160,100)(-120,-20)
\ArrowLine(0,100)(50,100)
\ArrowLine(50,100)(50,0)
\ArrowLine(50,0)(0,0)
\DashLine(50,100)(100,100){5}
\DashLine(50,0)(100,0){5}
\put(-10,-2){$\bar b$}
\put(-10,68){$b$}
\put(25,34){$t$}
\put(75,66){$H^-$}
\put(75,-4){$H^+$}
\put(31,-25){$(b)$}
\end{picture} \\[-0.5cm]
\setlength{\unitlength}{1pt}
\caption[ ]{\label{fg:dylodia} \it Typical diagrams contributing to
$q\bar q \to V^* \to H^+H^-$ at lowest order: (a) Drell--Yan-like
contribution, (b) bottom-Yukawa induced contribution.}
\end{figure}
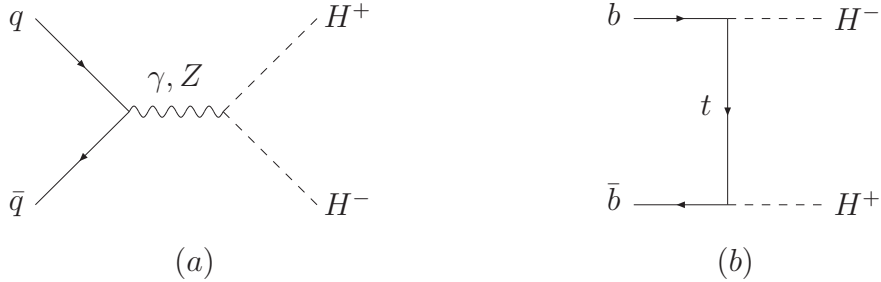
Within the type-I 2HDM the bottom Yukawa couplings are not enhanced so
that the contribution of bottom-Yukawa-induced processes, see
Fig.~\ref{fg:dylodia}b, starting from a $b\bar b$ state are tiny (at
the few per-mille level) compared to the leading Drell--Yan-like
contributions emerging from $s$-channel photon and $Z$-boson exchange,
see Fig.~\ref{fg:dylodia}a. Therefore, we are neglecting the
bottom-Yukawa-induced part in the following.  The lowest-order partonic
cross section can be expressed as \cite{dychar}
\beq
\hat\sigma_{\mathrm{LO}}(q\bar q\to H^+H^-)= \frac{\pi \alpha^2 \beta^3}{9 Q^2}
\left\{ e_q^2 + 2\frac{e_q v_q v_H}{1-\frac{M_Z^2}{Q^2}} +
\frac{(v_q^2+a_q^2) v_H}{\left( 1-\frac{M_Z^2}{Q^2}\right)^2} \right\}
\, ,
\label{eq:siglo}
\eeq
where $\beta = \sqrt{1-4 \frac{M_{H^\pm}^2}{Q^2}}$ denotes the usual two-body
phase-space factor and $v_q\, (a_q)$ are the (axial) vector couplings of the
quarks $q$
to the vector bosons $\gamma,Z$ and $v_H$ the one of the charged Higgs
bosons,
\beq
v_q = \frac{2I_{3q} - 4e_q s_W^2}{2s_Wc_W}\, , \qquad
a_q = \frac{2I_{3q}}{2s_Wc_W}\, , \qquad
v_H = \frac{1-2s_W^2}{2} \,,
\eeq
and $I_{3q} (e_q)$ denotes the third isospin component (electric charge)
of the quarks $q$. The sine of the Weinberg angle has been defined in
terms of the $W,Z$ masses, $s_W^2 = 1-M_W^2/M_Z^2$.
The partonic center-of-mass energy squared $\hat s$ coincides at
lowest order with the squared invariant mass $Q^2 = M^2_{H^+H^-}$ of the
charged Higgs pair, $\hat s=Q^2$. The hadronic cross section
can be obtained from convolving eq.~(\ref{eq:siglo}) with the corresponding
(anti)quark densities of the protons,
\beq
\sigma_{\mathrm{LO}}(pp \to H^+H^-) = \int_{\tau_0}^1 d\tau \sum_q
\frac{d{\cal L}^{q\bar q}}{d\tau} \hat\sigma_{\mathrm{LO}}(Q^2=\tau s) \, ,
\eeq
with $\tau_0 = 4M_{H^\pm}^2/s$ and $s$ the squared total hadronic
c.m.~energy.

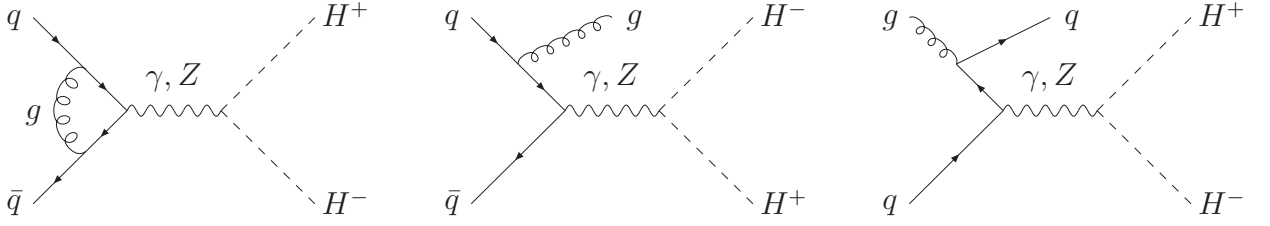
\begin{figure}[hbt]
\SetScale{0.7}
\begin{picture}(100,100)(-20,0)
\ArrowLine(0,100)(25,75)
\ArrowLine(25,75)(50,50)
\ArrowLine(50,50)(25,25)
\ArrowLine(25,25)(0,0)
\GlueArc(40,50)(25,118,241){4}{4}
\Photon(50,50)(100,50){3}{5}
\DashLine(100,50)(150,100){5}
\DashLine(100,50)(150,0){5}
\put(-10,-2){$\bar q$}
\put(-10,68){$q$}
\put(-3,32){$g$}
\put(42,45){$\gamma,Z$}
\put(108,67){$H^+$}
\put(108,-4){$H^-$}
\end{picture}
\begin{picture}(100,100)(-80,0)
\ArrowLine(0,100)(25,75)
\ArrowLine(25,75)(50,50)
\ArrowLine(50,50)(0,0)
\Gluon(25,75)(75,100){3}{5}
\Photon(50,50)(100,50){3}{5}
\DashLine(100,50)(150,100){5}
\DashLine(100,50)(150,0){5}
\put(-10,-2){$\bar q$}
\put(-10,68){$q$}
\put(58,68){$g$}
\put(42,45){$\gamma,Z$}
\put(108,67){$H^-$}
\put(108,-4){$H^+$}
\end{picture}
\begin{picture}(100,100)(-140,0)
\Gluon(0,100)(25,75){3}{3}
\ArrowLine(25,75)(75,100)
\ArrowLine(50,50)(25,75)
\ArrowLine(0,0)(50,50)
\Photon(50,50)(100,50){3}{5}
\DashLine(100,50)(150,100){5}
\DashLine(100,50)(150,0){5}
\put(-10,-2){$q$}
\put(-10,68){$g$}
\put(58,68){$q$}
\put(42,45){$\gamma,Z$}
\put(108,67){$H^+$}
\put(108,-4){$H^-$}
\end{picture} \\[-0.5cm]
\setlength{\unitlength}{1pt}
\caption[ ]{\label{fg:dynlodia} \it Typical diagrams contributing to
$q\bar q \to V^* \to H^+H^-$ at NLO.}
\end{figure}
The QCD corrections are identical to the corresponding corrections to
the Drell--Yan process, see Fig.~\ref{fg:dynlodia}. They modify the
lowest order cross section in the following way \cite{dynlo}:
\bea
\sigma(pp\to H^+H^-) & = & \sigma_{\mathrm{LO}} + \Delta\sigma_{q\bar q} + \Delta\sigma_{qg} \;,
\nonumber \\
\Delta\sigma_{q\bar q} & = & \frac{\alpha_s(\mu_R)}{\pi} \int_{\tau_0}^1 d\tau
\sum_q \frac{d{\cal L}^{q\bar q}}{d\tau} \int_{\tau_0/\tau}^1 dz~\hat
\sigma_{\mathrm{LO}}(Q^2 = \tau z s)~\omega_{q\bar q}(z) \nonumber\;, \\
\Delta\sigma_{qg} & = & \frac{\alpha_s(\mu_R)}{\pi} \int_{\tau_0}^1 d\tau
\sum_{q,\bar q} \frac{d{\cal L}^{qg}}{d\tau} \int_{\tau_0/\tau}^1 dz~\hat
\sigma_{\mathrm{LO}}(Q^2 = \tau z s)~\omega_{qg}(z)\;,
\eea
with the coefficient functions
\bea
\omega_{q\bar q}(z) & = & -P_{qq}(z) \log \frac{\mu_F^2}{\tau s}
+ \frac{4}{3}\left\{ 2[\zeta_2-2]\delta(1-z) + 4{\cal D}_1(z) - 2(1+z)\log(1-z)
\phantom{\frac{M^2}{s}}\!\!\!\!\!\!\!\!\!\!
\right\} \nonumber\,, \\
\omega_{qg}(z) & = & -\frac{1}{2} P_{qg}(z) \log \left(
\frac{\mu_F^2}{(1-z)^2 \tau s} \right) + \frac{1}{8}\left\{ 1+6z-7z^2 \right\} \, ,
\eea
where $\mu_R (\mu_F)$ denotes the renormalization (factorization) scale
and the Altarelli--Parisi splitting functions are given by \cite{altpar}
\bea
P_{qq}(z) & = & \frac{4}{3} \left\{ 2{\cal D}_0(z)-1-z+\frac{3}{2}\delta(1-z)
\right\} \nonumber \,, \\
P_{qg}(z) & = & \frac{1}{2} \left\{ z^2 + (1-z)^2 \right\} \, .
\label{eq:altpar}
\eea
The plus distributions ${\cal D}_i(z)$ involved in the expressions above read
\begin{equation}
{\cal D}_i(z) = \left( \frac{\log^i(1-z)}{1-z} \right)_+ \qquad
(i=0,1,\ldots) \; .
\label{eq:plus}
\end{equation}
The parton luminosity functions are given by
\begin{eqnarray}
\frac{d{\cal L}^{q\bar q}}{d\tau} & = & \int_\tau^1 \frac{dx}{x}~\left[
q(x,\mu_F^2) \bar q(\tau /x,\mu_F^2) + \bar q(x,\mu_F^2) q(\tau
/x,\mu_F^2) \right] \nonumber \,, \\
\frac{d{\cal L}^{qg}}{d\tau} & = & \int_\tau^1 \frac{dx}{x}~\left[ q(x,\mu_F^2)
g(\tau /x,\mu_F^2) + g(x,\mu_F^2) q(\tau /x,\mu_F^2) \right] \,.
\label{eq:partonlumi}
\end{eqnarray}
\begin{figure}[hbt]
\begin{center}
\begin{picture}(150,245)(0,0)
\put(-150,-155.0){\includegraphics{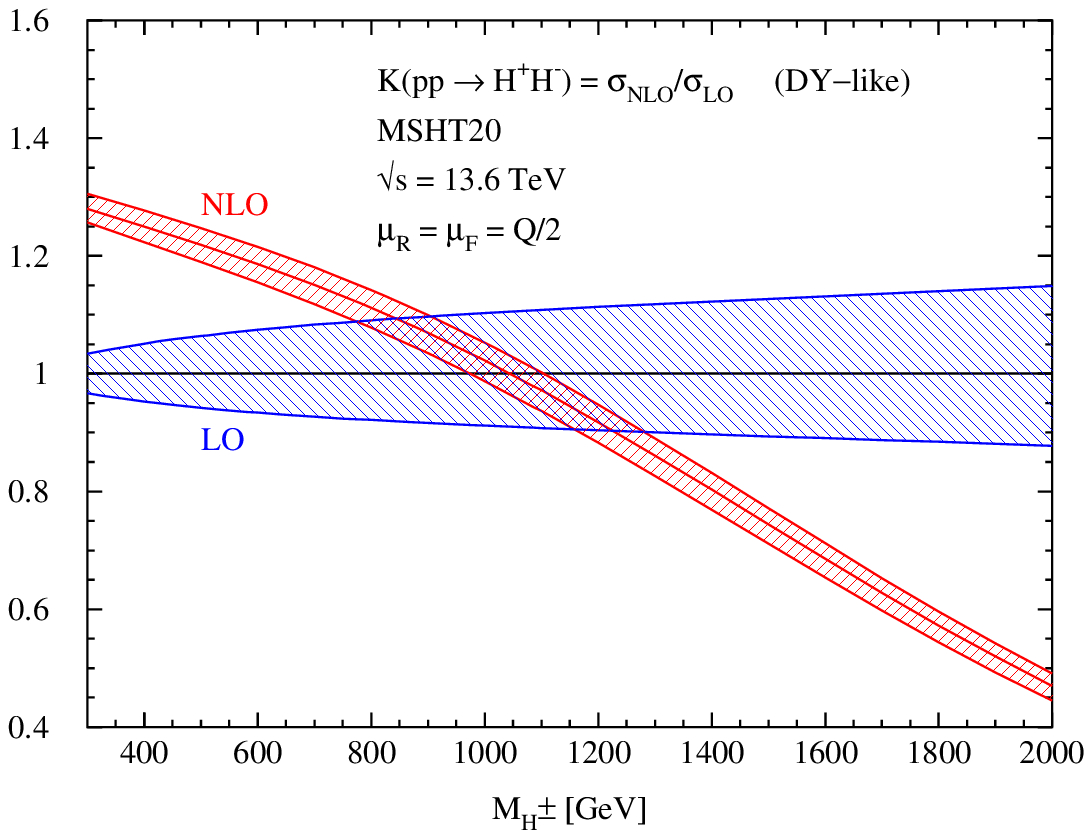}}
\end{picture}
\caption{\it The $K$ factor, defined as the ratio $\sigma_\mathrm{NLO}/\sigma_\mathrm{LO}$, for the Drell--Yan like cross section of
charged-Higgs pair production as a function of the charged Higgs mass with the corresponding uncertainty bands due to the scale dependence.
It is the same in both 2HDM scenarios. The MSHT20lo\_as130 and
MSHT20nlo\_as118 parton densities (PDFs) \cite{msht20} were used for the
LO and NLO cross sections, respectively, for a consistent definition of
the $K$ factor. The cross section is independent of $\tan\beta$ at LO
and NLO.}
\label{fg:k_dy}
\end{center}
\end{figure}
\begin{figure}[hbtp]
\begin{center}
\begin{picture}(150,400)(0,0)
\put(-150,-60.0){\includegraphics{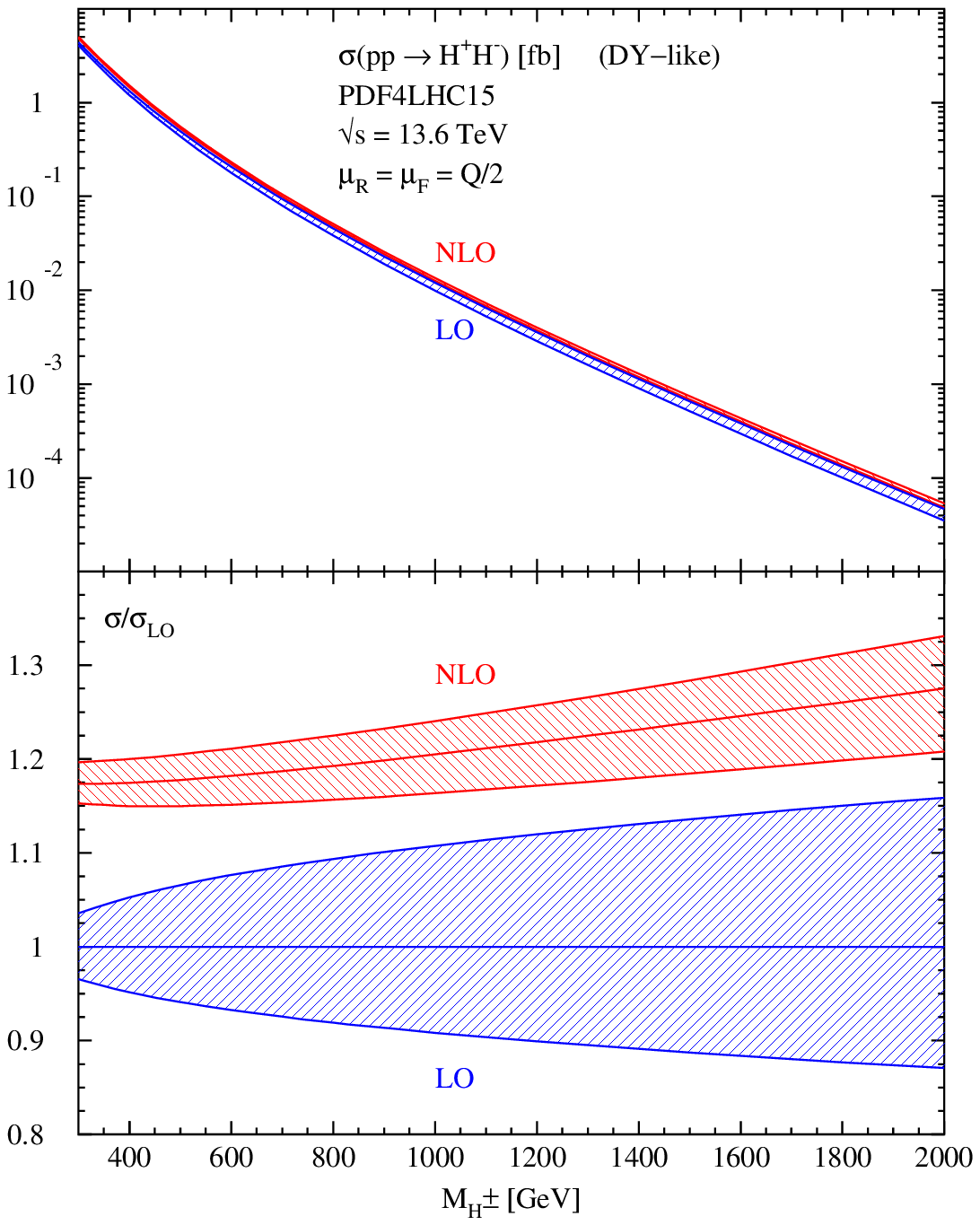}}
\end{picture}
\caption{\it Drell--Yan like cross section of charged-Higgs pair
production as a function of the charged Higgs mass with scale
uncertainties. The PDF4LHC15 parton densities \cite{pdf4lhc15} have
been used for the LO and NLO cross sections.}
\label{fg:sc_dy}
\end{center}
\end{figure}
\noindent
The QCD corrections to the Drell--Yan-like charged-Higgs pair production
are of more moderate size, i.e.~modifying the cross section by up to
about 30\%, when adopting the scale choice $\mu_R=\mu_F = Q/2$ (in line with recent scale choices for Higgs-pair production of the LHCHWG), where
$Q$ is the invariant mass of the charged Higgs pair, as is shown in
Fig.~\ref{fg:k_dy}\footnote{Fig.~\ref{fg:k_dy} includes the associated scale uncertainties at LO and NLO, too.}. 
However, for large charged Higgs masses the QCD
corrections decrease the cross section by about 50\%, but this effect
depends strongly on the NLO/LO PDF sets used in this calculation. Because the LO PDFs differ significantly between different PDF sets used in this calculation, there are PDF sets that lead to moderate corrections of about 10\% for large charged Higgs masses.
Since the cross section at leading order (LO) and NLO is independent of
$\tan\beta$, the $K$ factor and the related uncertainties are independent of
$\tan\beta$ as well. The scale uncertainties were determined by the
usual 7-point variation of the factorization and renormalization scales.
This results in residual generic theoretical uncertainties of up to
about 5\% at the NLO level with a significant reduction from the LO
prediction, see Fig.~\ref{fg:sc_dy}. The missing overlap between the LO
and NLO error bands indicates that the LO uncertainties are not valid in
quantitative terms. However, the significant reduction of the scale dependence from LO to NLO indicates the perturbative reliability of
the NLO predictions and its residual uncertainties that amount to
3--5\%. The Drell--Yan like charged-Higgs pair production process is
entirely mediated by gauge couplings and thus independent of $\tan\beta$
at LO and NLO.

\section{$H^+H^-$ Production in Gluon Fusion} \label{sc:ggf}
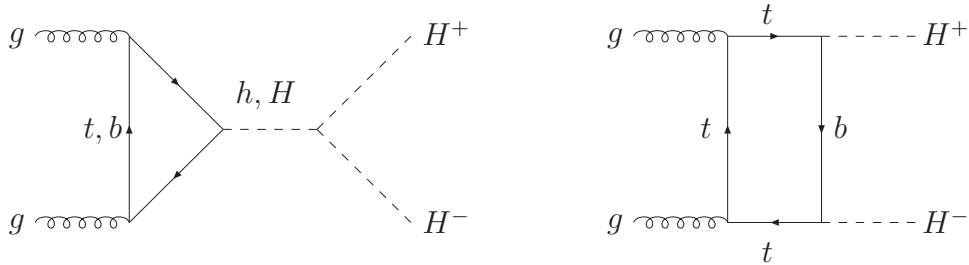
\begin{figure}[hbt]
\SetScale{0.7}
\begin{picture}(160,90)(-60,-20)
\Gluon(0,100)(50,100){3}{5}
\Gluon(0,0)(50,0){3}{5}
\ArrowLine(50,100)(100,50)
\ArrowLine(100,50)(50,0)
\ArrowLine(50,0)(50,100)
\DashLine(100,50)(150,50){5}
\DashLine(150,50)(200,100){5}
\DashLine(150,50)(200,0){5}
\put(-10,-2){$g$}
\put(-10,68){$g$}
\put(75,45){$h,H$}
\put(18,32){$t,b$}
\put(145,67){$H^+$}
\put(145,-4){$H^-$}
\end{picture}
\begin{picture}(160,90)(-120,-20)
\Gluon(0,100)(50,100){3}{5}
\Gluon(0,0)(50,0){3}{5}
\ArrowLine(50,100)(100,100)
\ArrowLine(100,100)(100,0)
\ArrowLine(100,0)(50,0)
\ArrowLine(50,0)(50,100)
\DashLine(100,100)(150,100){5}
\DashLine(100,0)(150,0){5}
\put(-10,-2){$g$}
\put(-10,68){$g$}
\put(48,75){$t$}
\put(48,-15){$t$}
\put(25,32){$t$}
\put(75,32){$b$}
\put(108,67){$H^+$}
\put(108,-4){$H^-$}
\end{picture} \\[-0.7cm]
\setlength{\unitlength}{1pt}
\caption[ ]{\label{fg:gghphmlodia} \it Typical diagrams contributing to
$gg \to H^+H^-$ at lowest order.}
\end{figure}
\noindent
Charged-Higgs pair production via gluon fusion is a loop-induced process
that is mediated by mixed top/bottom triangle and box diagrams at LO,
see Fig.~\ref{fg:gghphmlodia}. Generically the partonic LO cross section
can be expressed as \cite{gghphmlo}
\begin{equation}
\hat \sigma_{\mathrm{LO}}(gg\to H^+H^-)
 = \int_{\hat t_-}^{\hat t_+} d\hat t \,
\frac{G_F^2 \alpha_s^2(\mu_R)}{256 (2\pi)^3} \left\{ \left|
C_\triangle F_\triangle + F_\Box \right|^2 + \left| G_\Box \right|^2
+ \left| H_\Box \right|^2 \right\}.
\label{eq:gghphmlo}
\end{equation}
The Mandelstam variables for the parton process are given by
\begin{equation}
\hat s = Q^2, \qquad
\hat t/\hat u = -\frac{1}{2} 
\left[Q^2-2M_{H^\pm}^2 \mp Q^2 \beta \cos\theta\right], 
\end{equation}
where $Q^2$ denotes the squared invariant charged-Higgs pair mass, $Q^2
= M_{H^+H^-}^2$, and $\theta$ is the scattering angle in
the partonic center-of-mass system, while
\begin{equation}
\beta = \sqrt{ 1-4\frac{M_{H^\pm}^2}{Q^2}}\;.
\end{equation}
The integration limits
\begin{equation}
\hat t_\pm = -\frac{1}{2} \left[ Q^2 - 2M_{H^\pm}^2 \mp Q^2 \beta \right]
\end{equation}
in Eq.\ (\ref{eq:gghphmlo}), correspond to $\cos\theta=\pm 1$. The scale
parameter $\mu_R$ $(\mu_F)$ is the renormalization (factorization)
scale. The complete dependence on the fermion masses is contained in the
functions $F_\triangle$, $F_\Box$, $G_\Box$ and $H_\Box$. The full
expressions of the form factors $F_\triangle$, $F_\Box$, $G_\Box$,
$H_\Box$, including the exact dependence on the fermion masses, can be
found in Ref.~\cite{gghphmlo}.

The coupling $C_\triangle$ in general, and the form factors $F_\triangle,
F_\Box, G_\Box, H_\Box$ in the HTL are given by
\begin{eqnarray}
C_\triangle & = & \sum_{H_i = h,H} \frac{\lambda_{H_iH^+H^-}}
{\hat s - M_{H_i}^2}\, , 
\qquad
F_\triangle \to \frac{2}{3}~g_t^{H_i} \, , \quad 
F_\Box \to \frac{2}{3\tan^2\beta}\, , \quad
G_\Box, H_\Box \to 0\, ,
\end{eqnarray}
where the trilinear scalar Higgs couplings to charged Higgs states are given by
\bea
\lambda_{hH^+H^-} & = & (M_h^2+2M_{H^\pm}^2-2M^2) s_{\beta-\alpha} + 2
c_{\beta-\alpha} \cot(2\beta) (M_h^2-M^2) \,,\\
\lambda_{HH^+H^-} & = & (M_H^2+2M_{H^\pm}^2-2M^2) c_{\beta-\alpha} - 2
s_{\beta-\alpha} \cot(2\beta) (M_H^2-M^2) \, .
\eea
and the neutral Higgs Yukawa-coupling factors read
\bea
g_t^h = \frac{s_\alpha}{s_\beta} \, , \qquad\qquad\qquad g_t^H = \frac{c_\alpha}{s_\beta} \, .
\eea
The expressions in the HTL can be obtained from the effective Lagrangian
\beq
{\cal L}_{\mathrm{eff}} = \frac{\alpha_s}{24\pi} G^{a\mu\nu} G^a_{\mu\nu}~\log
\left( \frac{2|\Phi_2|^2}{v_2^2} \right)~\left( 1+\frac{11}{4}
\frac{\alpha_s}{\pi} \right) \, ,
\eeq
including NLO QCD corrections\footnote{The NLO QCD corrections to the
effective Lagrangian are the same for the charged and neutral states.},
where $G^{a\mu\nu}$ denotes the gluon field strength tensor, and the
second Higgs doublet $\Phi_2$ of Eq.~(\ref{eq:vevexpansion}) is used.
Expanding the Higgs fields of $\Phi_2$ according to
Eq.~(\ref{eq:diagGHpm}), one arrives at the relevant part of the effective
Lagrangian for this process
\beq
{\cal L}_{\mathrm{eff}} = \frac{\alpha_s}{12\pi v} G^{a\mu\nu} G^a_{\mu\nu}~\left\{
h \frac{s_\alpha}{s_\beta} + H \frac{c_\alpha}{s_\beta} + \frac{H^+
H^-}{v \tan^2 \beta} +\cdots \right\}~\left( 1+\frac{11}{4} \frac{\alpha_s}{\pi}
\right) \, .
\label{eq:leff}
\eeq
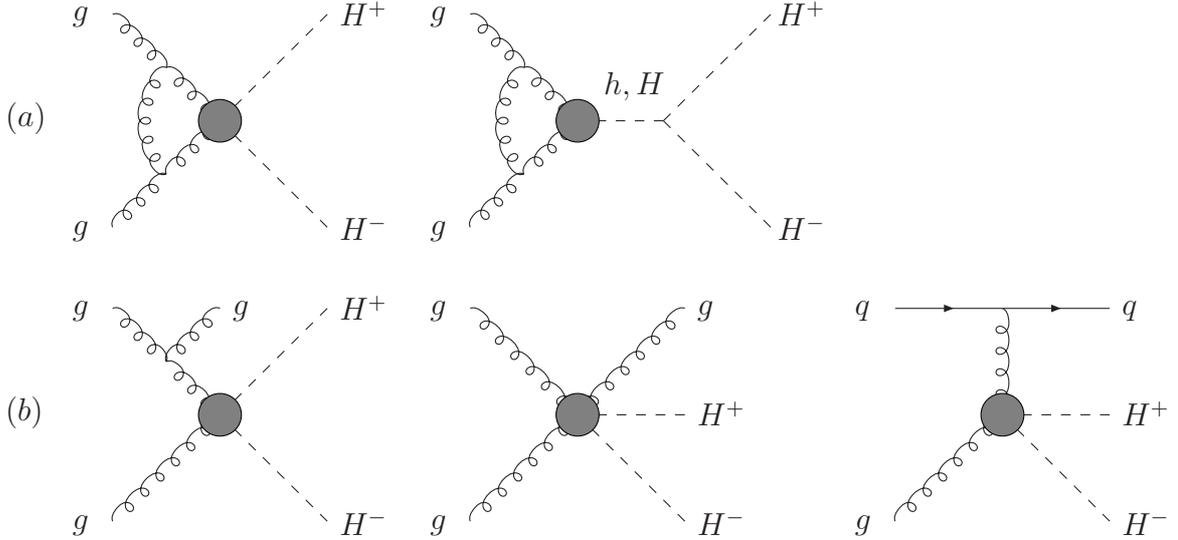
\begin{figure}[t]
\begin{picture}(100,90)(-50,-5)
\SetScale{0.8}
\Gluon(0,100)(25,75){3}{3}
\Gluon(25,75)(50,50){3}{3}
\Gluon(0,0)(25,25){3}{3}
\Gluon(25,25)(50,50){3}{3}
\GlueArc(50,50)(35.355,135,225){3}{5}
\DashLine(50,50)(100,100){5}
\DashLine(50,50)(100,0){5}
\GCirc(50,50){10}{0.5}
\put(85,76){$H^+$}
\put(85,-5){$H^-$}
\put(-15,-3){$g$}
\put(-15,78){$g$}
\put(-40,38){$(a)$}
\SetScale{1}
\end{picture}
\begin{picture}(100,90)(-80,-5)
\SetScale{0.8}
\Gluon(0,100)(25,75){3}{3}
\Gluon(25,75)(50,50){3}{3}
\Gluon(0,0)(25,25){3}{3}
\Gluon(25,25)(50,50){3}{3}
\GlueArc(50,50)(35.355,135,225){3}{5}
\DashLine(50,50)(90,50){5}
\DashLine(90,50)(140,100){5}
\DashLine(90,50)(140,0){5}
\GCirc(50,50){10}{0.5}
\put(115,76){$H^+$}
\put(115,-5){$H^-$}
\put(50,50){$h,H$}
\put(-15,-3){$g$}
\put(-15,78){$g$}
\SetScale{1}
\end{picture} \\
\begin{picture}(100,110)(-50,-5)
\SetScale{0.8}
\Gluon(0,100)(25,75){3}{3}
\Gluon(25,75)(50,50){3}{3}
\Gluon(25,75)(50,100){3}{3}
\Gluon(0,0)(50,50){3}{6}
\DashLine(50,50)(100,100){5}
\DashLine(50,50)(100,0){5}
\GCirc(50,50){10}{0.5}
\put(85,76){$H^+$}
\put(85,-5){$H^-$}
\put(-15,-3){$g$}
\put(-15,78){$g$}
\put(45,78){$g$}
\put(-40,38){$(b)$}
\SetScale{1}
\end{picture}
\begin{picture}(100,110)(-80,-5)
\SetScale{0.8}
\Gluon(0,100)(50,50){3}{6}
\Gluon(0,0)(50,50){3}{6}
\Gluon(50,50)(100,100){3}{6}
\DashLine(50,50)(100,50){5}
\DashLine(50,50)(100,0){5}
\GCirc(50,50){10}{0.5}
\put(85,36){$H^+$}
\put(85,-5){$H^-$}
\put(-15,-3){$g$}
\put(-15,78){$g$}
\put(85,78){$g$}
\SetScale{1}
\end{picture}
\begin{picture}(100,110)(-135,-5)
\SetScale{0.8}
\ArrowLine(0,100)(50,100)
\ArrowLine(50,100)(100,100)
\Gluon(50,100)(50,50){3}{4}
\Gluon(0,0)(50,50){3}{6}
\DashLine(50,50)(100,50){5}
\DashLine(50,50)(100,0){5}
\GCirc(50,50){10}{0.5}
\put(85,36){$H^+$}
\put(85,-5){$H^-$}
\put(-15,-3){$g$}
\put(-15,78){$q$}
\put(85,78){$q$}
\SetScale{1}
\end{picture} \\
\caption[]{\it \label{fg:gghphmnlodia} Typical effective diagrams contributing
to the (a) virtual and (b) real corrections to charged Higgs-boson pair 
production via gluon fusion, where for the blobs, the effective Feynman
rules derived from the Lagrangian in Eq.~(\ref{eq:leff}) are used.}
\end{figure}
\noindent
Using this effective Lagrangian, the NLO QCD corrections can be
approximated by the ones obtained in the HTL (see
Fig.~\ref{fg:gghphmnlodia}), but convolved with the full LO partonic
cross section kernel of Eq.~(\ref{eq:gghphmlo}),
\begin{equation}
\sigma_{\mathrm{NLO}}(pp \rightarrow H^+ H^- + X) = 
\sigma_{\mathrm{LO}} + \Delta
\sigma_{\mathrm{virt}} + \Delta\sigma_{gg} + \Delta\sigma_{gq} + \Delta\sigma_{q\bar{q}}\;,
\label{eq:gghqcd}
\end{equation}
with the individual contributions
\begin{eqnarray}
\sigma_{\mathrm{LO}} & = & \int_{\tau_0}^1 d\tau~\frac{d{\cal L}^{gg}}{d\tau}~
\hat\sigma_{\mathrm{LO}}(Q^2 = \tau s)\;, 
\nonumber \\ 
\Delta \sigma_{\mathrm{virt}} & = & \frac{\alpha_s(\mu_R)} {\pi}\int_{\tau_0}^1 d\tau~
\frac{d{\cal L}^{gg}}{d\tau}~\hat \sigma_{\mathrm{LO}}(Q^2=\tau
s)~\left\{ \pi^2 + \frac{11}{2} + \frac{33-2N_F}{6} \log
\frac{\mu_R^2}{Q^2} \right\}\;, 
\nonumber \\ 
\Delta \sigma_{gg} & = & \frac{\alpha_{s}(\mu_R)} {\pi} \int_{\tau_0}^1 d\tau~
\frac{d{\cal L}^{gg}}{d\tau} \int_{\tau_0/\tau}^1 \frac{dz}{z}~
\hat\sigma_{\mathrm{LO}}(Q^2 = z \tau s)
\left\{ - z P_{gg} (z) \log \frac{\mu_F^{2}}{\tau s} \right. 
\nonumber \\
& & \left. \hspace*{3.0cm} {} - \frac{11}{2} (1-z)^3 + 6 [1+z^4+(1-z)^4]\,
{\cal D}_1(z) \right\}, 
\nonumber \\ 
\Delta \sigma_{gq} & = & \frac{\alpha_{s}(\mu_R)} {\pi} \int_{\tau_0}^1 d\tau
\sum_{q,\bar{q}} \frac{d{\cal L}^{gq}}{d\tau} \int_{\tau_0/\tau}^1 \frac{dz}{z}~
\hat \sigma_{\mathrm{LO}}(Q^2 = z \tau s)
\left\{ -\frac{z}{2} P_{gq}(z) \log\frac{\mu_F^{2}}{\tau s(1-z)^2} \right. 
\nonumber \\
& & \left. \hspace*{9.0cm} {} + \frac{2}{3}z^2 - (1-z)^2 
\vphantom{\frac{M^{2}}{\tau s(1-z)^2}} \right\},
\nonumber \\ 
\Delta \sigma_{q\bar q} & = & \frac{\alpha_s(\mu_R)}
{\pi} \int_{\tau_0}^1 d\tau
\sum_{q} \frac{d{\cal L}^{q\bar q}}{d\tau} \int_{\tau_0/\tau}^1 \frac{dz}{z}~
\hat \sigma_{\mathrm{LO}}(Q^2 = z \tau s)~\frac{32}{27} (1-z)^3 \, ,
\end{eqnarray}
where
\begin{equation}
\tau_0 = 4\frac{M_{H^\pm}^2}{s}\; .
\end{equation}
The objects $P_{gg}(z), P_{gq}(z)$ denote the
Altarelli--Parisi splitting functions \cite{altpar}
\begin{eqnarray}
P_{gg}(z) &=& 6\left\{ {\cal D}_0(z)
+\frac{1}{z}-2+z(1-z) \right\} + \frac{33-2N_F}{6}\delta(1-z)\,, 
\nonumber \\
P_{gq}(z) &=& \frac{4}{3} \frac{1+(1-z)^2}{z}\,,
\end{eqnarray}
with $N_F=5$ in our case. The parton--parton luminosities
$d{\cal L}^{ij}/d\tau$ are given in Eq.~(\ref{eq:partonlumi}) and
\begin{eqnarray}
\frac{d{\cal L}^{gg}}{d\tau} & = & \int_\tau^1 \frac{dx}{x}~g(x,\mu_F^2)
g(\tau /x,\mu_F^2) \, .
\end{eqnarray}
The results above are adopted from the corresponding calculation in the
single-Higgs case \cite{gghnlo} and neutral Higgs-boson pair production
\cite{gghhnlohtl} in the HTL. Compared to the neutral-Higgs pair
production there is no one-particle reducible contribution with gluon
exchange in the $t$-channel between two of the effective vertices here,
so that the virtual corrections are simpler.

\begin{figure}[hbt]
\begin{center}
\begin{picture}(150,250)(0,0)
\put(-150,-155.0){\includegraphics{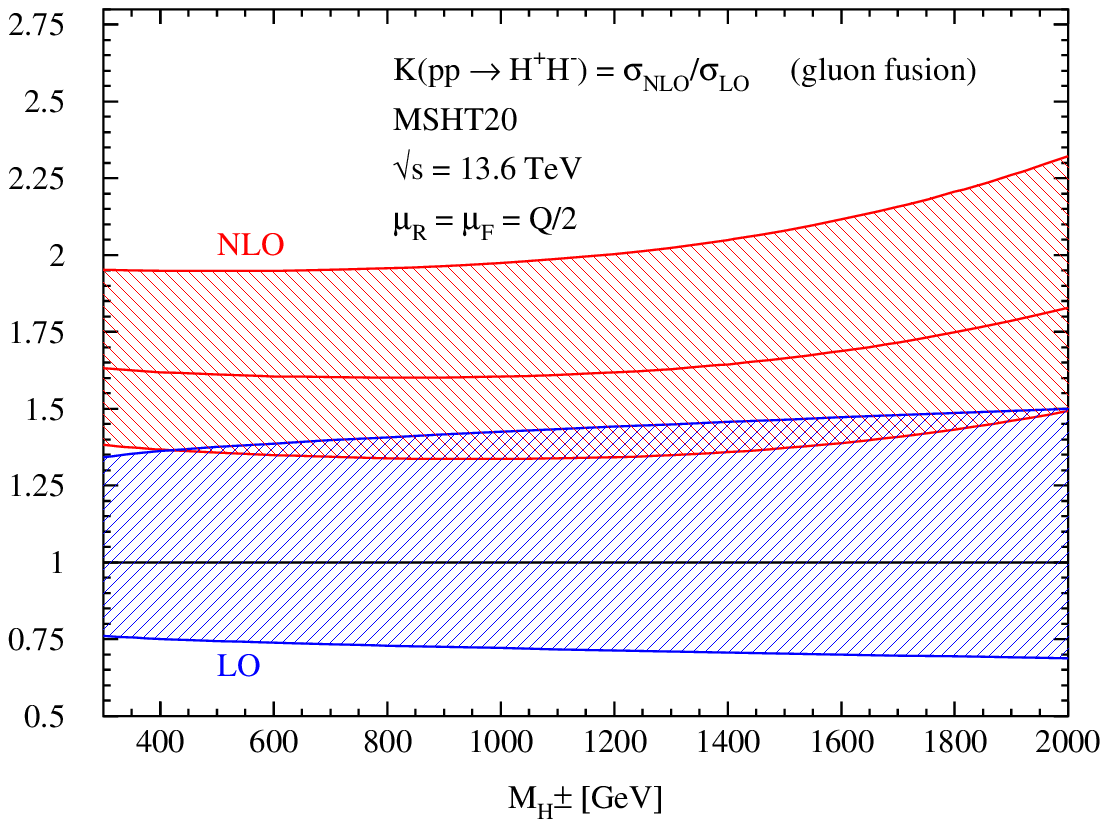}}
\end{picture}
\caption{\it The $K$ factor of the gluon-fusion cross section describing
charged-Higgs pair production as a function of the charged Higgs mass with the corresponding uncertainty bands due to the scale dependence
for the two 2HDM scenarios, where the central K factor for $\tan\beta=2$ is shown in red, while the results for $\tan\beta=20$ are hardly distinguishable and thus not shown. The MSHT20lo\_as130 and MSHT20nlo\_as118
PDFs \cite{msht20} were adopted for the LO and NLO cross sections,
respectively, for the consistent definition of the $K$ factor.} 
\label{fg:k_ggf}
\end{center}
\end{figure}
\begin{figure}[hbtp]
\begin{center}
\begin{picture}(150,400)(0,0)
\put(-150,-60.0){\includegraphics{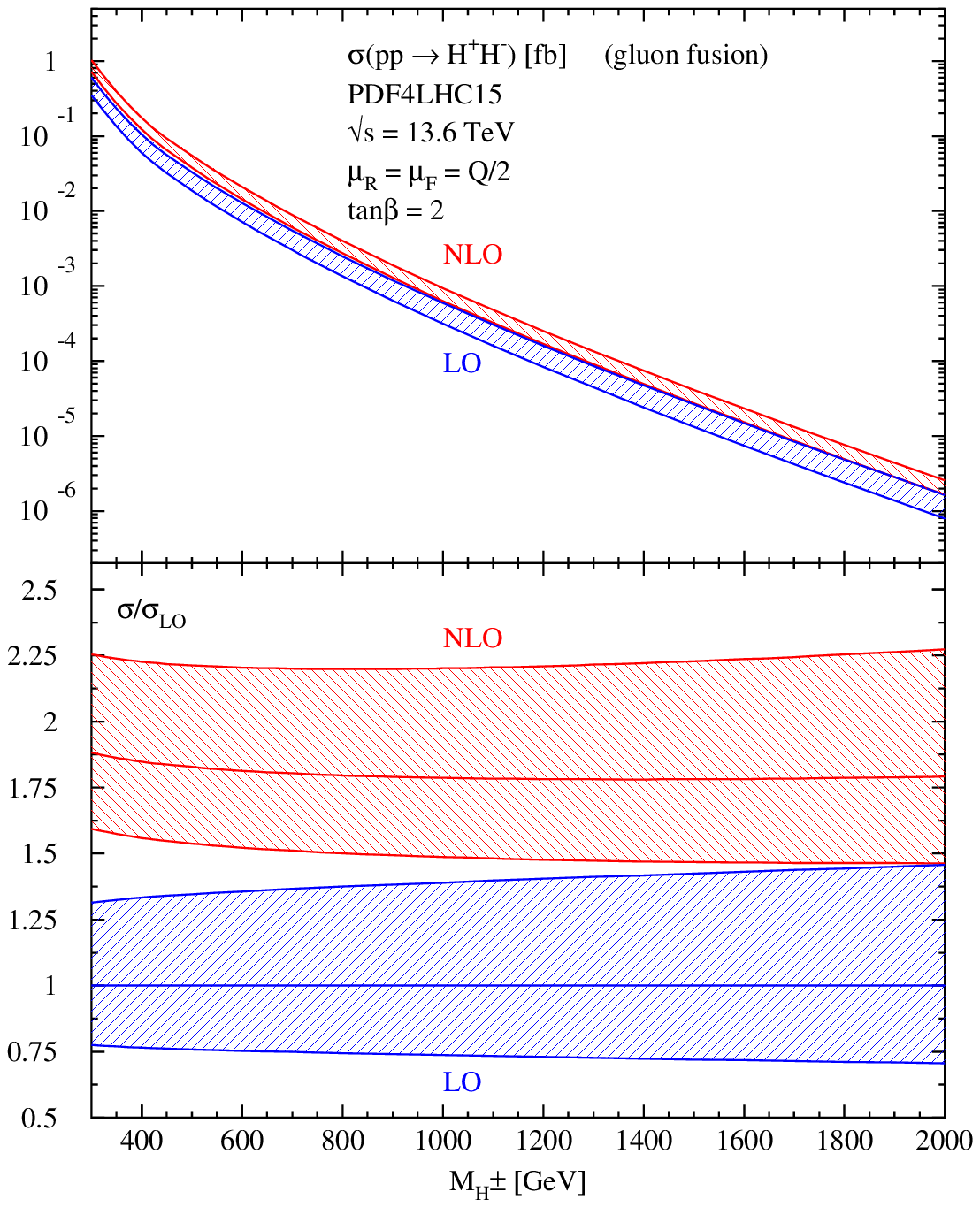}}
\end{picture}
\caption{\it Gluon-fusion cross section of charged-Higgs pair production
as a function of the charged Higgs mass with scale uncertainties for the
scenario with $\tan\beta = 2$. The PDF4LHC15 PDFs \cite{pdf4lhc15} were
used for the LO and NLO cross sections.}
\label{fg:sc_ggf}
\end{center}
\end{figure}
\begin{figure}[hbtp]
\begin{center}
\begin{picture}(150,400)(0,0)
\put(-150,-60.0){\includegraphics{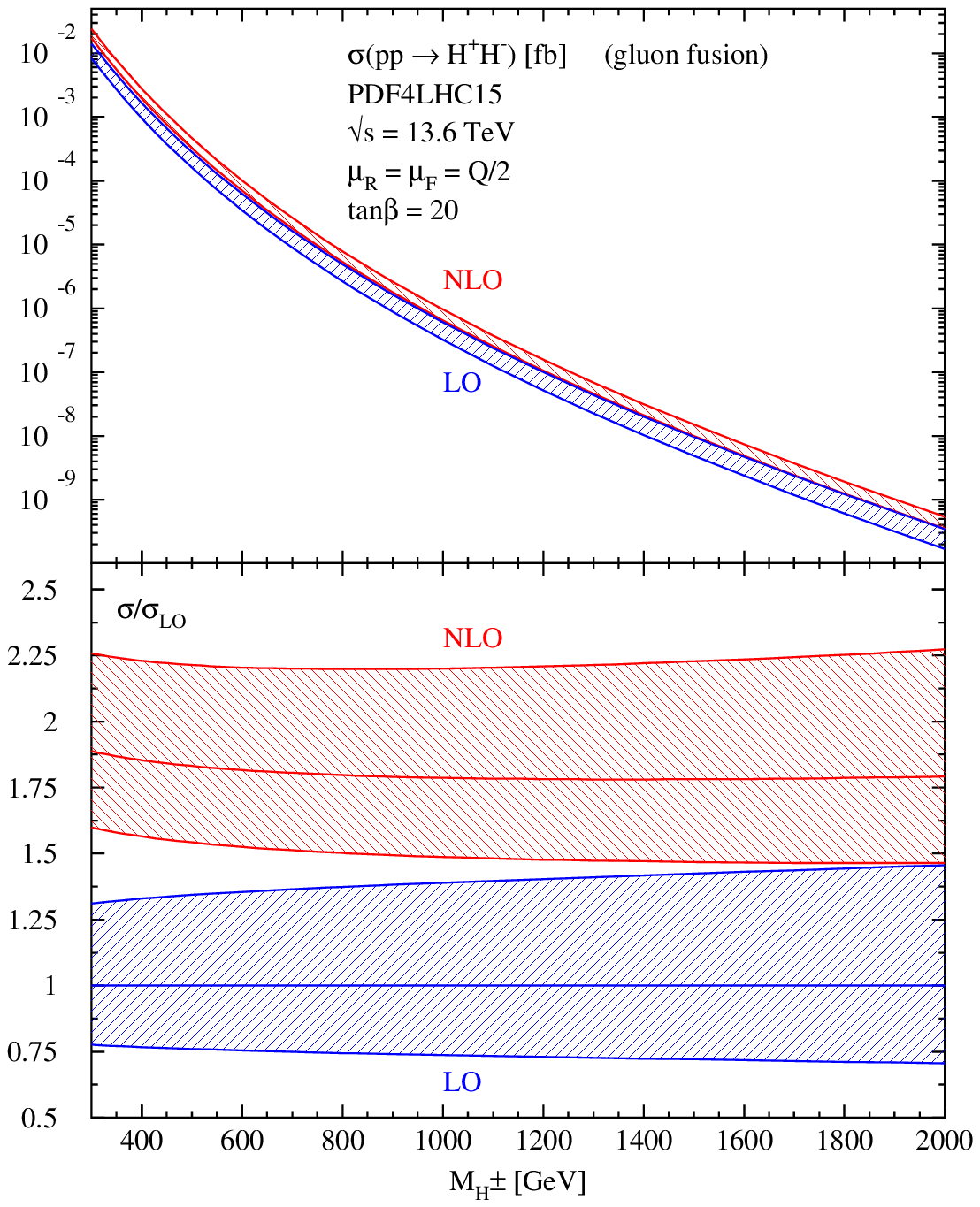}}
\end{picture}
\caption{\it The same as Fig.~\ref{fg:sc_ggf} but for $\tan\beta = 20$.}
\label{fg:sc_ggf1}
\end{center}
\end{figure}

The QCD corrections to the gluon-fusion cross section are large\footnote{We find reasonable agreement with the approximate results of Ref.~\cite{Hespel:2014sla}.}
i.e.~increasing the cross section by up to about 80\% for the central
scale choice of half the invariant charged-Higgs pair mass, see
Fig.~\ref{fg:k_ggf}\footnote{Fig.~\ref{fg:k_ggf} includes the corresponding LO and NLO scale-uncertainty bands as well. The central K factors are hardly different between $\tan\beta = 2,20$.}. Thus their inclusion is highly relevant to arrive
at a reliable prediction of the production rate\footnote{The
neglected quark-mass effects at NLO in the HTL are expected at the 20--30\%
level in line with the observations for the neutral-Higgs pair
production via gluon fusion \cite{gghhnlo}.}. The relative QCD corrections
develop only a minor dependence on the value of $\tan\beta$, while the
inclusive cross section exhibits a strong $\tan\beta$ dependence, see
Figs.~\ref{fg:sc_ggf}, \ref{fg:sc_ggf1}. The residual renormalization
and factorization scale dependences were determined by the
7-point method. The residual uncertainties at NLO appear to be at the
level of up to 20--25\% for the scale choice of half the invariant
charged-Higgs pair mass, see Figs.~\ref{fg:sc_ggf}, \ref{fg:sc_ggf1}.
Also in this case we do not observe an overlap between the LO and NLO scale variation bands so that the LO uncertainty estimate is not reliable.
This is in line with the observations in neutral-Higgs pair
production via gluon fusion \cite{gghhnlohtl}.

\section{$H^+H^-$ Production in Vector-Boson Fusion} \label{sc:vbf}
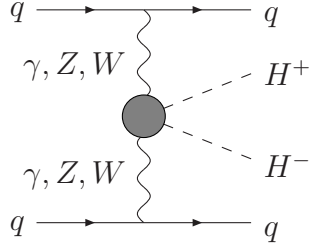
\begin{figure}[hbt]
\begin{center}
\begin{picture}(100,100)(0,-10)
\SetScale{0.8}
\ArrowLine(0,100)(50,100)
\ArrowLine(50,100)(100,100)
\ArrowLine(0,0)(50,0)
\ArrowLine(50,0)(100,0)
\Photon(50,100)(50,50){3}{3}
\Photon(50,50)(50,0){3}{3}
\DashLine(50,50)(100,70){5}
\DashLine(50,50)(100,30){5}
\GCirc(50,50){10}{0.5}
\put(-5,56){$\gamma,Z,W$}
\put(-5,15){$\gamma,Z,W$}
\put(85,52){$H^+$}
\put(85,17){$H^-$}
\put(85,76){$q$}
\put(85,-5){$q$}
\put(-10,-3){$q$}
\put(-10,78){$q$}
\SetScale{1}
\end{picture}
\caption[]{\it \label{fg:vbfhphmdia} Generic diagram of vector-boson
fusion at LO. The blob contains all subdiagrams of the processes
$V_1V_2\to H^+H^-$ with $V_{1,2} = \gamma,Z,W$.}
\end{center}
\end{figure}

Charged-Higgs pairs can also be produced via VBF,
where the Higgs pair emerges from photons, $Z$ bosons or $W$
bosons emitted from the incoming quarks and antiquarks, see
Fig.~\ref{fg:vbfhphmdia}.
We are treating this process in the structure-function approach (SFA) \cite{sfa} that
neglects interference terms in the case of identical incoming quarks
that arise between diagrams of the type of Fig.~\ref{fg:vbfhphmdia}
with the corresponding one with crossed quark lines in the final state.
In this approach, the squared matrix element of the process can be
expressed as
\bea
|{\cal M}|^2 & = & 4q_1^2 q_2^2 \sum_{V_{1,\ldots,4} = \gamma,Z}
\frac{g_{V_1} g_{V_2} g_{V_3} g_{V_4}}{(q_1^2-M_{V_1}^2)
(q_2^2-M_{V_2}^2) (q_1^2-M_{V_3}^2) (q_2^2-M_{V_4}^2)} W^{\mu\rho}_{V_1
V_3} W^{\nu\sigma}_{V_2V_4} {\cal M}^{\mu\nu}_{V_1V_2} {\cal
M}^{\ast\rho\sigma}_{V_3V_4} \nonumber \\
& + & \frac{4q_1^2 q_2^2}{(q_1^2-M_W^2)^2 (q_2^2-M_W^2)^2} \sum_{WW' =
W^+W^-, W^-W^+}
W^{\mu\rho}_{WW} W^{\nu\sigma}_{W'W'} {\cal M}^{\mu\nu}_{WW'} {\cal
M}^{\ast\rho\sigma}_{WW'} \, ,
\label{eq:vbf}
\eea
where $q_{1,2}^2$ denote the squared momenta of the individual vector
bosons at the corresponding proton leg and the gauge-coupling factors
are given by
\beq
g_\gamma = e, \qquad g_Z = \frac{M_Z}{2v}, \qquad g_W = \frac{M_W}{2v}
\, .
\eeq
\begin{figure}[hbt]
\SetScale{0.7}
\begin{picture}(100,100)(-40,-5)
\Photon(0,100)(50,100){5}{4}
\Photon(0,0)(50,0){5}{4}
\DashLine(50,100)(50,0){5}
\DashLine(50,100)(100,100){5}
\DashLine(50,0)(100,0){5}
\put(-25,-2){$\gamma/Z$}
\put(-25,68){$\gamma/Z$}
\put(40,33){$H^\pm$}
\put(75,66){$H^+$}
\put(75,-4){$H^-$}
\end{picture}
\begin{picture}(100,100)(-80,-5)
\Photon(0,100)(50,50){5}{5}
\Photon(0,0)(50,50){5}{5}
\DashLine(50,50)(100,100){5}
\DashLine(50,50)(100,0){5}
\put(-25,-2){$\gamma/Z$}
\put(-25,68){$\gamma/Z$}
\put(75,66){$H^+$}
\put(75,-4){$H^-$}
\end{picture}
\begin{picture}(100,100)(-120,-5)
\Photon(0,100)(50,50){5}{5}
\Photon(0,0)(50,50){5}{5}
\DashLine(50,50)(100,50){5}
\DashLine(100,50)(150,100){5}
\DashLine(100,50)(150,0){5}
\put(-15,-2){$Z$}
\put(-15,68){$Z$}
\put(45,45){$h,H$}
\put(112,66){$H^+$}
\put(112,-4){$H^-$}
\end{picture} \\
\begin{picture}(100,100)(-40,-5)
\Photon(0,100)(50,100){5}{4}
\Photon(0,0)(50,0){5}{4}
\DashLine(50,100)(50,0){5}
\DashLine(50,100)(100,100){5}
\DashLine(50,0)(100,0){5}
\put(-25,-2){$W^-$}
\put(-25,68){$W^+$}
\put(40,33){$h,H,A$}
\put(75,66){$H^+$}
\put(75,-4){$H^-$}
\end{picture}
\begin{picture}(100,100)(-80,-5)
\Photon(0,100)(50,50){5}{5}
\Photon(0,0)(50,50){5}{5}
\DashLine(50,50)(100,100){5}
\DashLine(50,50)(100,0){5}
\put(-25,-2){$W^-$}
\put(-25,68){$W^+$}
\put(75,66){$H^+$}
\put(75,-4){$H^-$}
\end{picture}
\begin{picture}(100,100)(-120,-5)
\Photon(0,100)(50,50){5}{5}
\Photon(0,0)(50,50){5}{5}
\DashLine(50,50)(100,50){5}
\DashLine(100,50)(150,100){5}
\DashLine(100,50)(150,0){5}
\put(-25,-2){$W^-$}
\put(-25,68){$W^+$}
\put(45,45){$h,H$}
\put(112,66){$H^+$}
\put(112,-4){$H^-$}
\end{picture} \\
\begin{picture}(100,100)(-180,-5)
\Photon(0,100)(50,50){5}{5}
\Photon(0,0)(50,50){5}{5}
\Photon(50,50)(100,50){5}{4}
\DashLine(100,50)(150,100){5}
\DashLine(100,50)(150,0){5}
\put(-25,-2){$W^-$}
\put(-25,68){$W^+$}
\put(45,45){$\gamma,Z$}
\put(112,66){$H^+$}
\put(112,-4){$H^-$}
\end{picture} \\[-0.7cm]
\setlength{\unitlength}{1pt}
\caption[ ]{\label{fg:subdia} \it Typical diagrams contributing to
the subprocesses of charged-Higgs pair production via VBF.}
\end{figure}
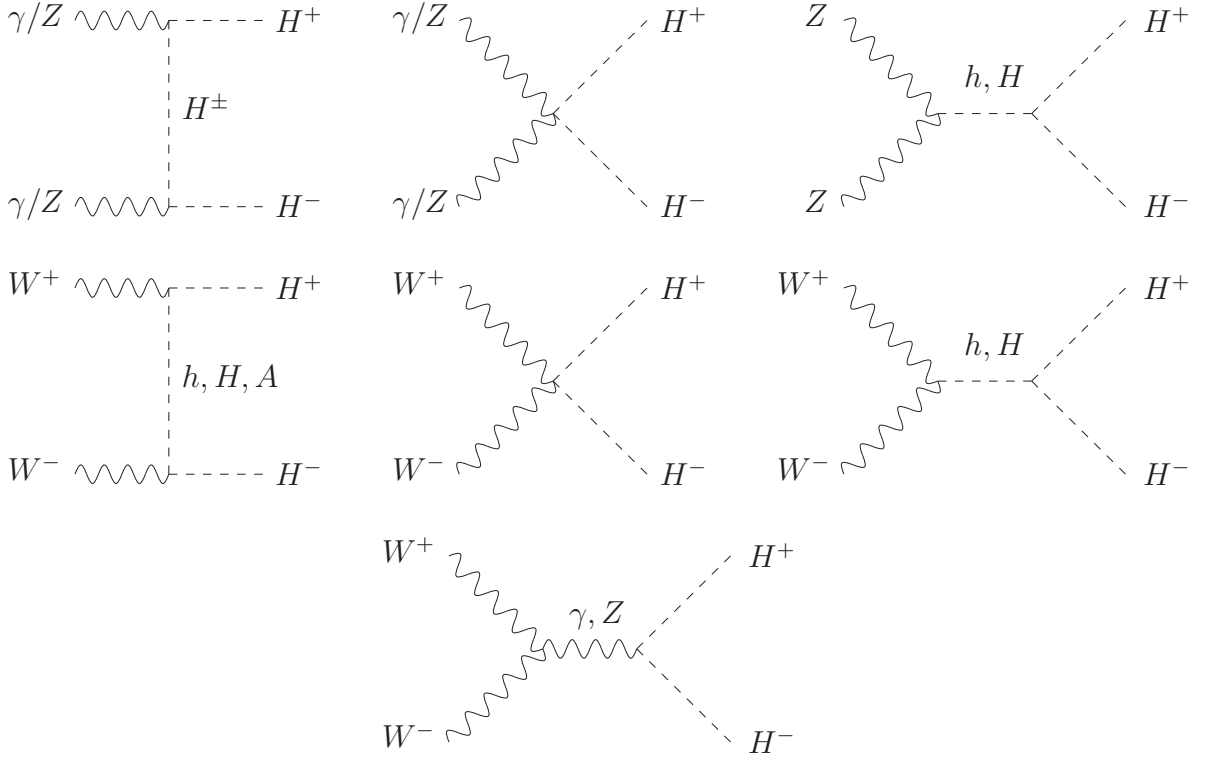
\noindent
The matrix elements of the subprocesses $V_1(q_1) V_2(q_2)
\to H^+(p_1) H^-(p_2)$ ($V_{1,2} = \gamma, Z, W$), see
Fig.~\ref{fg:subdia}, can be cast into the form
\bea
{\cal M}^{\mu\nu}_{W^+W^-} & = & i g_W^2 \left\{
\sum_{\phi=h,H,A} \frac{\left| \tilde g_W^\phi\right|^2}{\tilde t -
M_\phi^2} (q_1-2p_1)^\mu (q_2-2p_2)^\nu
+ 2g^{\mu\nu} \left[
1+\sum_{H_i=h,H} \frac{g_W^{H_i} \lambda_{H_iH^+H^-}}{\tilde s -
M_{H_i}^2} \right] \right\} \nonumber \\
& - & \!\!\! \sum_{V=\gamma,Z} i\frac{g_{3V} g_V c_V}{\tilde s - M_V^2} \left\{
g^{\mu\nu} (p_1-p_2)(q_1-q_2) - (p_1-p_2)^\mu (q_1+q)^\nu +
(p_1-p_2)^\nu (q_2+q)^\mu \right\} \nonumber \\
{\cal M}^{\mu\nu}_{V_1V_2} & = & i g_{V_1} g_{V_2} \left\{ c_{V_1} c_{V_2}
\left[ \frac{(q_1-2p_1)^\mu (q_2-2p_2)^\nu}{\tilde t - M_{H^\pm}^2} +
\frac{(q_1-2p_2)^\mu (q_2-2p_1)^\nu}{\tilde u - M_{H^\pm}^2} + 2
g^{\mu\nu} \right] \right. \nonumber \\
& & \left. \hspace*{2cm} + \delta_{V_1 Z} \delta_{V_2 Z} 2g^{\mu\nu}
\sum_{H_i=h,H} g_V^{H_i} \frac{\lambda_{H_iH^+H^-}}{\tilde s -
M_{H_i}^2} \right\} \qquad (V_{1,2} = \gamma,Z)
\eea
with $c_\gamma = 1$, $c_Z = 1-2s_W^2$, $\tilde s =
(q_1+q_2)^2$, $\tilde t = (p_1-q_1)^2$ and $\tilde u = (p_1-q_2)^2$. The
other coupling factors are given by
\beq
g_{W/Z}^h = -\tilde g_W^H = s_{\beta-\alpha}\, , \quad g_{W/Z}^H = \tilde g_W^h = c_{\beta-\alpha}
\, , \quad \tilde g_W^A = -i \, , \quad
g_{3V} = \left\{ \begin{array}{ll} e & ; V=\gamma \\
\displaystyle 2\frac{M_W}{v} c_W & ; V=Z \end{array} \right. \, .
\eeq
The contributing hadronic tensors are defined in terms of the usual
structure functions $F_{1\ldots 3}$ of deep inelastic lepton-nucleon
scattering (where we need to distinguish between the corresponding
vector-bosons)
\beq
W^{\mu\nu}_{V_1V_2} = F_{1,V_1V_2} \left( -g^{\mu\nu} + \frac{q^\mu q^\nu}{q^2} \right)
- \frac{F_{2,V_1V_2}}{xpq} \left( p^\mu - \frac{pq}{q^2} q^\mu \right)
\left( p^\nu - \frac{pq}{q^2} q^\nu \right)
- i \frac{F_{3,V_1V_2}}{2pq} \epsilon^{\mu\nu\rho\sigma} p_\rho q_\sigma
\, .
\eeq
The 4-momentum $p$ denotes the incoming {\it quark} momentum and $q$ the
{\it outgoing} gauge-boson momentum. The sums of Eq.~(\ref{eq:vbf}) run
over all contributing combinations of photon and $Z$ exchanges as well
as charged $W$ exchanges (The neutral and charged vector-boson exchanges
do not interfere.). The individual structure functions are defined
as\footnote{For charged $W$-exchange contributions we do not take into
account bottom PDFs, since they would lead to top quarks in the final
state.}
\begin{eqnarray}
F_{1,V_1V_2}(x,\mu_F^2) & = & \sum_q (v_{q,V_1} v_{q,V_2}+a_{q,V_1} a_{q,V_2}) [q(x,\mu_F^2) + \bar q(x,\mu_F^2)] \nonumber \,, \\
F_{2,V_1V_2}(x,\mu_F^2) & = & 2x \sum_q (v_{q,V_1} v_{q,V_2}+a_{q,V_1} a_{q,V_2}) [q(x,\mu_F^2) + \bar q(x,\mu_F^2)] \nonumber \,,\\
F_{3,V_1V_2}(x,\mu_F^2) & = & 2 \sum_q (v_{q,V_1} a_{q,V_2}+a_{q,V_1}
v_{q,V_2}) [-q(x,\mu_F^2) + \bar q(x,\mu_F^2)] \, ,
\label{eq:stfu}
\end{eqnarray}
where the electroweak coupling factors are given by
\bea
\gamma: & v_{q,\gamma} = e_q \, , \;\;\,\qquad\qquad & \qquad a_{q,\gamma} = 0 \,, \nonumber \\
Z: & v_{q,Z} = 2I_{3q} - 4e_q s_W^2 \, , & \qquad a_{q,Z} = 2I_{3q} \,,\nonumber \\
W: & v_{q,W} = \sqrt{2} \, , \;\qquad\qquad & \qquad a_{q,W} = \sqrt{2}
\, .
\eea
Finally, the differential cross section reads
\beq
d\sigma = \frac{\sum |{\cal M}|^2}{8 s} \frac{dx_1 dx_2}{x_1 x_2} dPS_4\,,
\eeq
with $s$ denoting the squared hadronic c.m.~energy and $x_1,x_2$ the two
Bjorken-$x$ values of the individual proton legs. The last factor
$dPS_4$ is the suitably parametrized 4-particle phase space measure at the
parton level. In line with the usual treatment of VBF
in single-Higgs and neutral-Higgs pair production we are
introducing a lower cut on the virtualities of the vector bosons, $Q_i^2
= -q_i^2 > 4$ GeV$^2$ ($i=1,2$).

\begin{figure}[hbtp]
\begin{center}
\begin{picture}(150,260)(0,0)
\put(-150,-155.0){\includegraphics{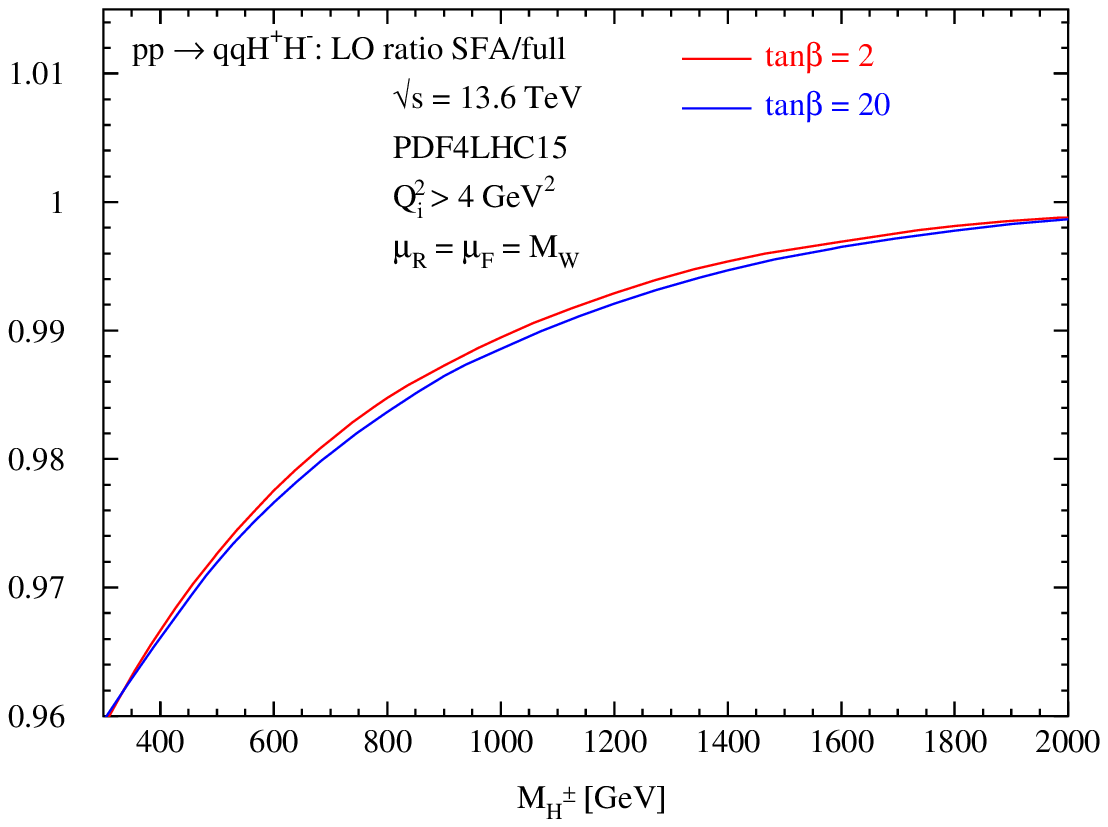}}
\end{picture}
\caption{\it The ratio between the SFA and the full calculation of charged-Higgs pair production via VBF at LO for the two 2HDM scenarios
of Eqs.~(\ref{eq:scenario1},\ref{eq:scenario2}) as a function of the
charged Higgs mass. The PDF4LHC15 PDFs \cite{pdf4lhc15} were
used for the LO cross sections.}
\label{fg:ratio}
\end{center}
\end{figure}
For the two scenarios above, we have checked the validity of the SFA by
comparing with the full calculation at LO. The LO calculation of the
matrix element has been performed with SUSY--MadGraph \cite{smadgraph}
by omitting all non-VBF diagrams \cite{simone}\footnote{Our comparison with the former LO calculation of Ref.~\cite{moretti} revealed several sources of discrepancies related to sign errors and incorrect definitions of electroweak couplings.} The ratios between the
SFA and the full calculation at LO are shown in Fig.~\ref{fg:ratio} for a fixed scale choice $\mu_R = \mu_F=M_W$. We
observe that the SFA agrees with the full calculation within about 3\% for charged Higgs masses above about 400 GeV.

\begin{figure}[hbtp]
\begin{center}
\begin{picture}(150,250)(0,0)
\put(-150,-155.0){\includegraphics{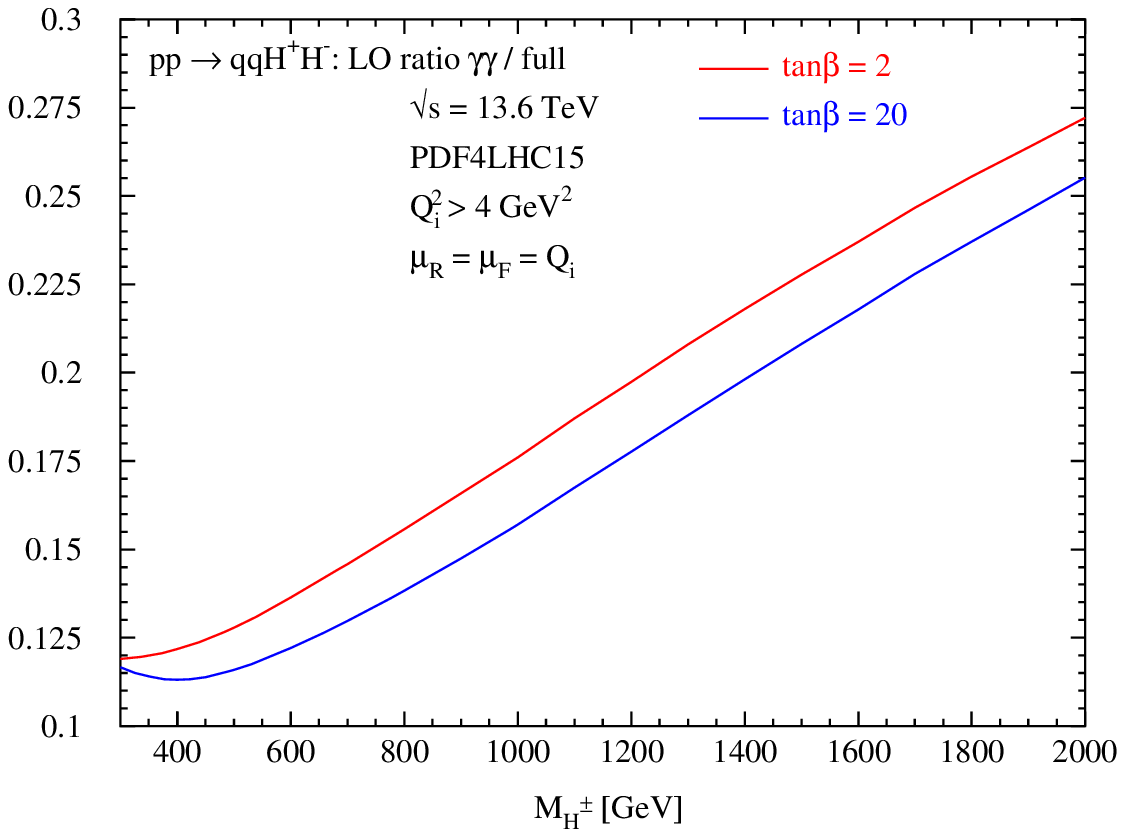}}
\end{picture}
\caption{\it The ratio between the pure photon-exchange contribution and the
full calculation within the SFA of charged-Higgs pair production via
VBF at LO for the two 2HDM scenarios of
Eqs.~(\ref{eq:scenario1},\ref{eq:scenario2}) as a function of the charged
Higgs mass. The PDF4LHC15 PDFs \cite{pdf4lhc15} were used for the LO
cross sections.}
\label{fg:gamma_full}
\end{center}
\end{figure}

Very recently, a study has appeared on charged-Higgs pair production via
photon fusion at the LHC \cite{gagahphm}, i.e.~neglecting $Z$ and $W$ exchange
diagrams. This approximation is compared with our full calculation at LO in
Fig.~\ref{fg:gamma_full}, where the ratio of both calculations is depicted. It
is clearly visible that photon-fusion only makes up less than 30\% of the full
results of charged-Higgs pair production. However, we did not include the
experimental cuts the authors of Ref.~\cite{gagahphm} have applied, but the
full cross section is significantly larger than just photon fusion so that the
analysis of Ref.~\cite{gagahphm} requires more theoretical refinements.

Within the structure-function approach the NLO QCD corrections can be
implemented by shifting the structure functions by their known radiative
corrections
\begin{eqnarray}
F_{i,V_1V_2}(x,\mu_F^2) & \to & F_{i,V_1V_2}(x,\mu_F^2) + \Delta F_{i,V_1V_2}(x,\mu_F^2,Q^2) \hspace*{1cm} (i=1,2,3)\,,
\end{eqnarray}
and expanding the full matrix element to NLO. The explicit NLO
correction terms read
\begin{eqnarray}
\Delta F_{1,V_1V_2}(x,\mu_F^2,Q^2) & = & \frac{\alpha_s(\mu_R)}{\pi}\sum_q
(v_{q,V_1} v_{q,V_2} + a_{q,V_1} a_{q,V_2})
\int_x^1 \frac{dy}{y} \left\{ \frac{2}{3} [q(y,\mu_F^2) + \bar q(y,\mu_F^2)]
\right. \nonumber \\
& &
\left[ -\frac{3}{4} P_{qq}(z) \log \frac{\mu_F^2z}{Q^2} + (1+z^2) {\cal D}_1(z)
- \frac{3}{2} {\cal D}_0(z) \right. \nonumber \\
& & \left. \hspace*{6cm} + 3 - \left(
\frac{9}{2} + \frac{\pi^2}{3} \right) \delta(1-z) \right] 
\nonumber \\
& & \left. + \frac{1}{4} g(y,\mu_F^2) \left[ -2 P_{qg}(z) \log
\frac{\mu_F^2z}{Q^2(1-z)} + 4z(1-z)
- 1 \right] \right\} \,, \nonumber \\
\Delta F_{2,V_1V_2}(x,\mu_F^2,Q^2) & = & 2x\frac{\alpha_s(\mu_R)}{\pi}\sum_q
(v_{q,V_1} v_{q,V_2} + a_{q,V_1} a_{q,V_2})
\int_x^1 \frac{dy}{y} \left\{ \frac{2}{3} [q(y,\mu_F^2) + \bar q(y,\mu_F^2)]
\right. \nonumber \\
& &
\left[ -\frac{3}{4} P_{qq}(z) \log \frac{\mu_F^2z}{Q^2} + (1+z^2) {\cal D}_1(z)
- \frac{3}{2} {\cal D}_0(z) \right. \nonumber \\
& & \left. \hspace*{3.0cm} + 3 + 2z - \left(
\frac{9}{2} + \frac{\pi^2}{3} \right) \delta(1-z) \right]
\nonumber \\
& & \left. + \frac{1}{4} g(y,\mu_F^2) \left[ -2P_{qg}(z) \log
\frac{\mu_F^2z}{Q^2(1-z)}
+ 8z(1-z) - 1 \right] \right\} \nonumber \,, \\
\Delta F_{3,V_1V_2}(x,\mu_F^2,Q^2) & = & \frac{\alpha_s(\mu_R)}{\pi} \sum_q 2
(v_{q,V_1} a_{q,V_2} + a_{q,V_1} v_{q,V_2})
\int_x^1 \frac{dy}{y} \left\{ \frac{2}{3} [-q(y,\mu_F^2) + \bar
q(y,\mu_F^2)]
\right. \nonumber \\
& &
\left[ -\frac{3}{4} P_{qq}(z) \log \frac{\mu_F^2z}{Q^2} + (1+z^2) {\cal D}_1(z)
- \frac{3}{2} {\cal D}_0(z) \right. \nonumber \\
& & \left. \left. \hspace*{3cm} + 2 + z - \left(
\frac{9}{2} + \frac{\pi^2}{3} \right) \delta(1-z) \right] \right\} \, ,
\end{eqnarray}
where $z=x/y$ and the plus distributions of Eq.~(\ref{eq:plus}) have
been used. The Altarelli--Parisi splitting functions $P_{qq}$ and
$P_{qg}$ are defined in Eq.~(\ref{eq:altpar}).

\begin{figure}[hbt]
\begin{center}
\begin{picture}(150,245)(0,0)
\put(-150,-155.0){\includegraphics{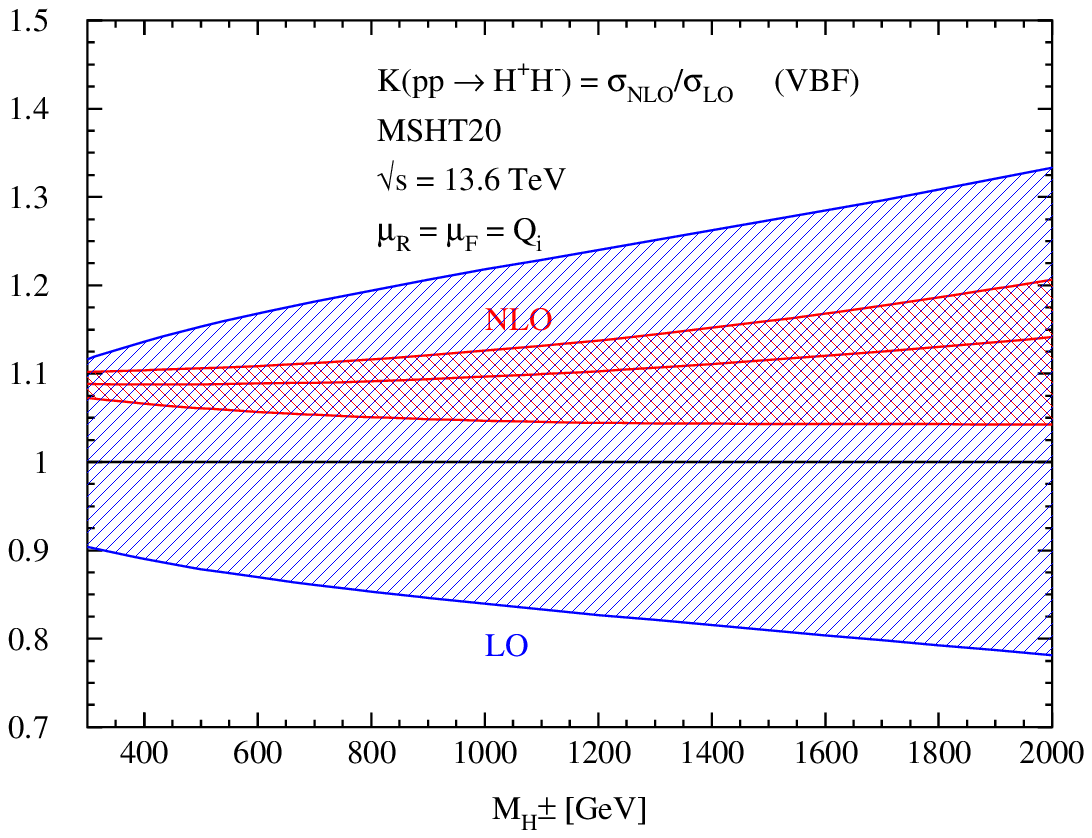}}
\end{picture}
\caption{\it The $K$ factor for the VBF cross section of
charged-Higgs pair production as a function of the charged Higgs mass with the corresponding uncertainty bands due to the scale dependence
for the two 2HDM scenarios, where the central K factor for $\tan\beta=2$ is shown in red, while the results for $\tan\beta=20$ are hardly distinguishable and thus not shown. The MSHT20lo\_as130 and MSHT20nlo\_as118
PDFs \cite{msht20} were adopted for the LO and NLO cross sections,
respectively.}
\label{fg:k_vbf}
\end{center}
\end{figure}
\begin{figure}[hbtp]
\begin{center}
\begin{picture}(150,400)(0,0)
\put(-150,-60.0){\includegraphics{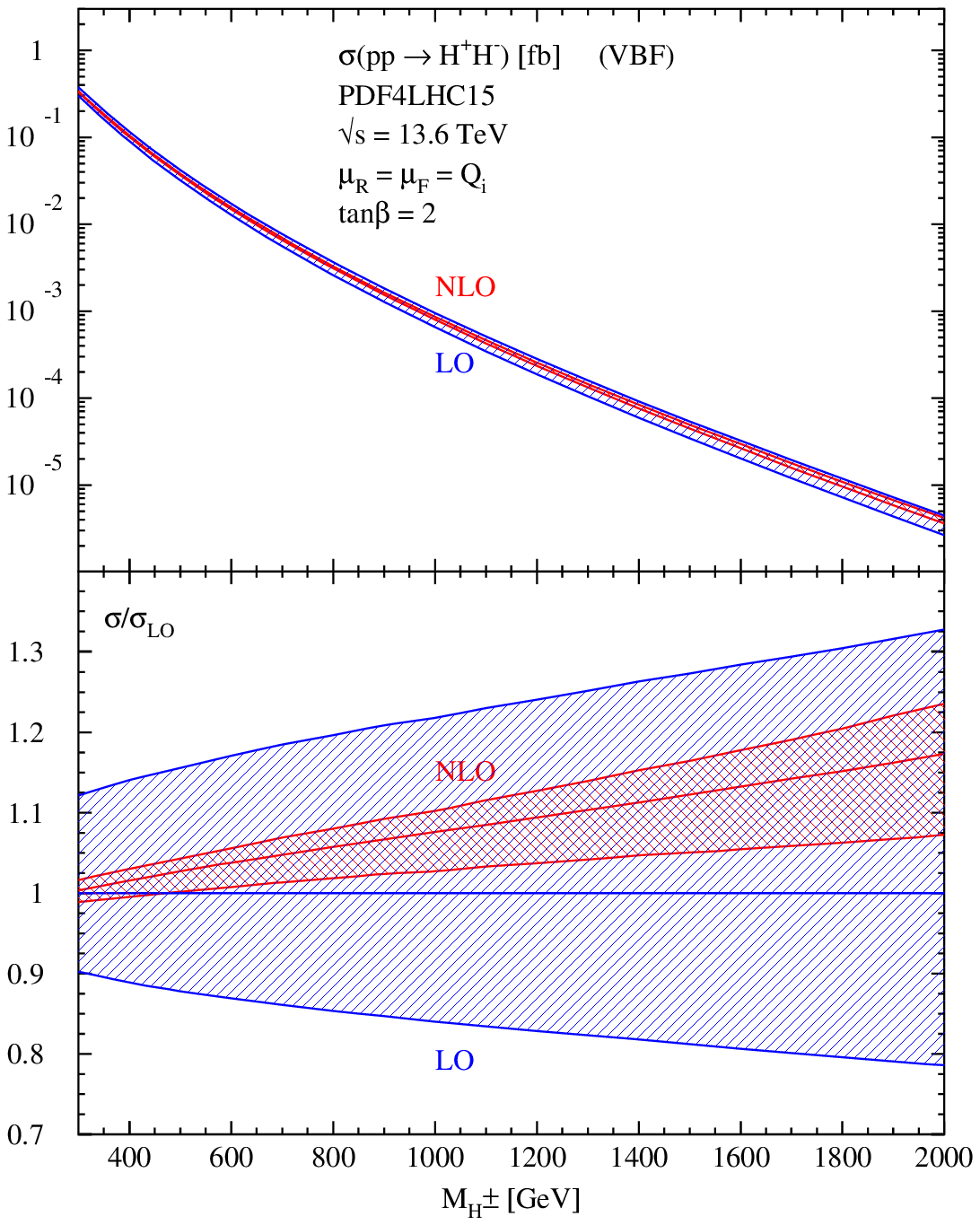}}
\end{picture}
\caption{\it The cross section of VBF of
charged-Higgs pair production as a function of the charged Higgs mass
for $\tan\beta=2$. The PDF4LHC15 PDFs \cite{pdf4lhc15} were adopted for the LO and NLO cross sections.}
\label{fg:sc_vbf1}
\end{center}
\end{figure}
\begin{figure}[hbtp]
\begin{center}
\begin{picture}(150,400)(0,0)
\put(-150,-60.0){\includegraphics{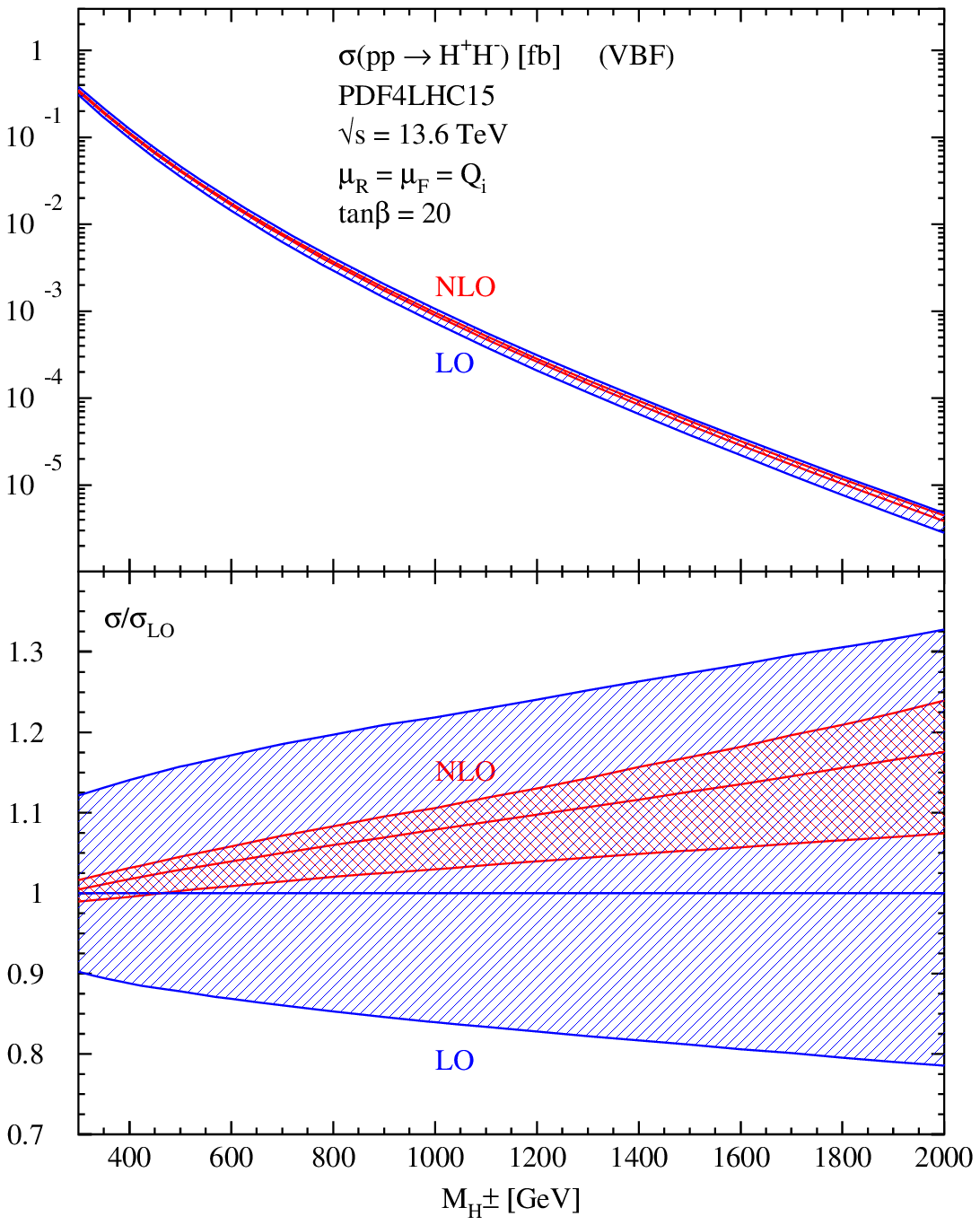}}
\end{picture}
\caption{\it The same as Fig.~\ref{fg:sc_vbf1} but for $\tan\beta = 20$.}
\label{fg:sc_vbf2}
\end{center}
\end{figure}

\begin{figure}[hbtp]
\begin{center}
\begin{picture}(150,245)(0,0)
\put(-150,-155.0){\includegraphics{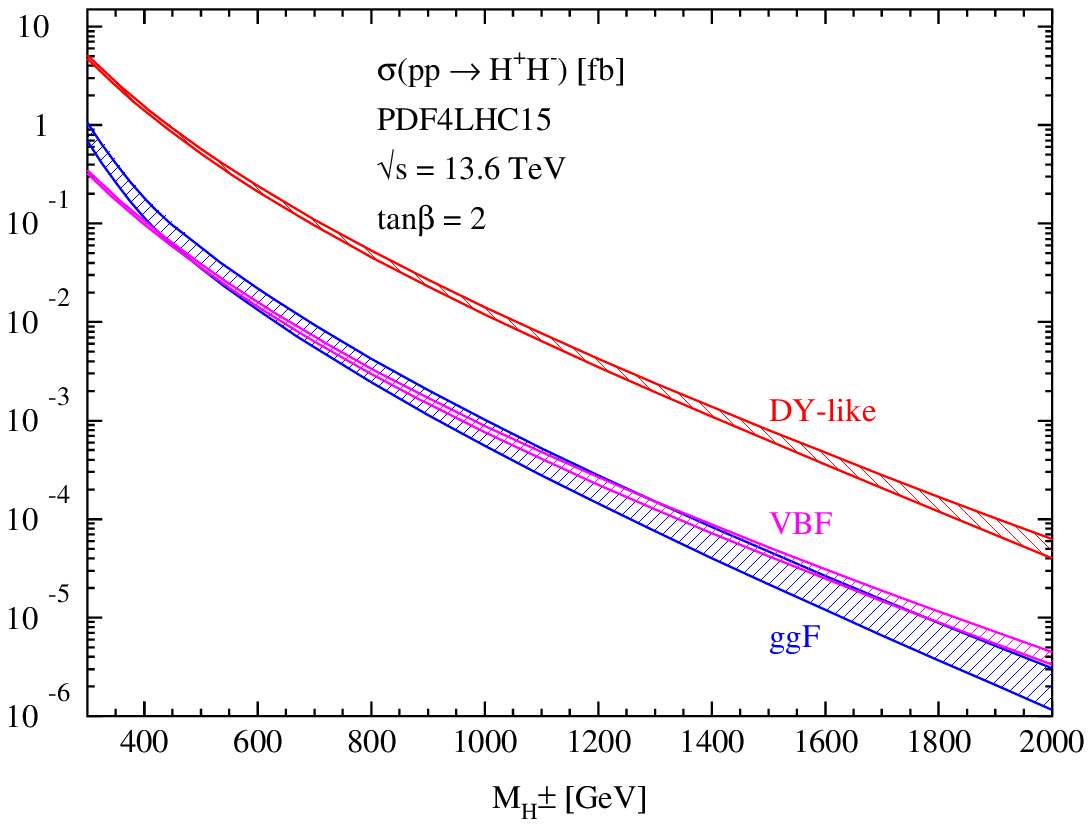}}
\end{picture}
\caption{\it The total cross sections of all production modes of charged-Higgs pair production as a function of the charged Higgs mass for
$\tan\beta = 2$. The PDF4LHC15 PDFs \cite{pdf4lhc15} were adopted for
the NLO cross sections. The error bands are the sum of the scale and the
PDF+$\alpha_s$ uncertainties.}
\label{fg:cxn_2}
\end{center}
\end{figure}
\begin{figure}[hbtp]
\begin{center}
\begin{picture}(150,230)(0,0)
\put(-150,-155.0){\includegraphics{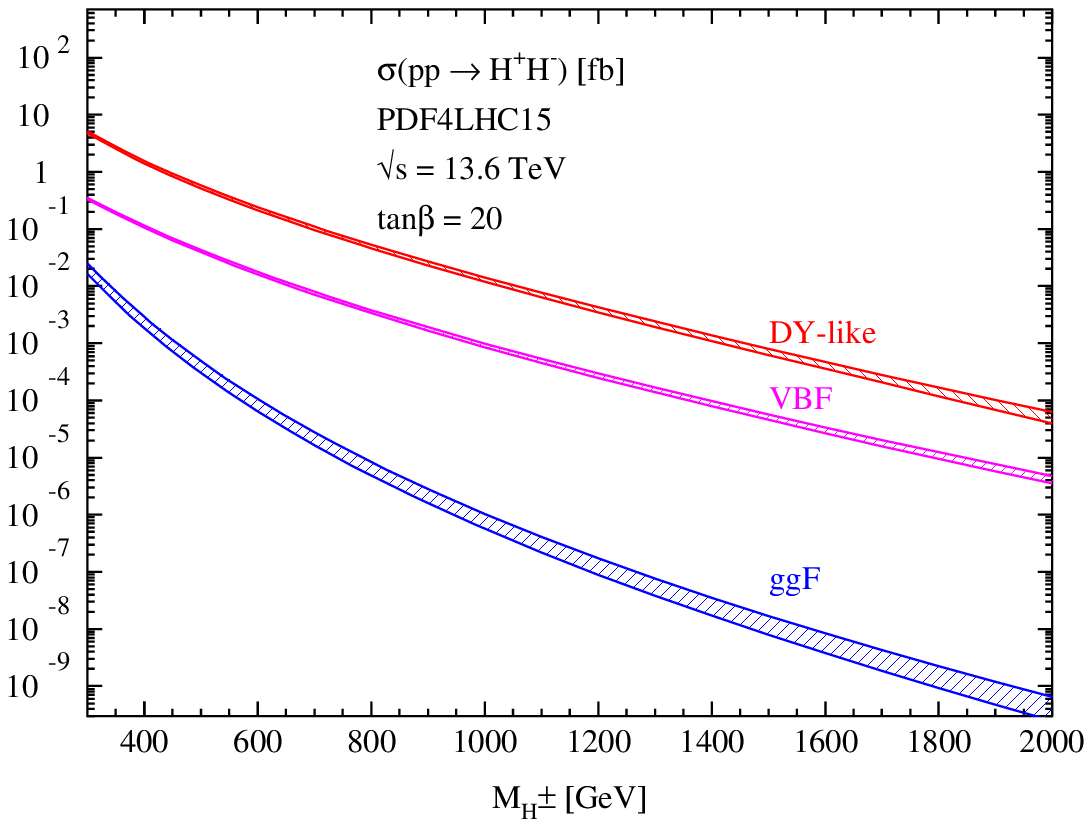}}
\end{picture}
\caption{\it The same as Fig.~\ref{fg:cxn_2} but for $\tan\beta = 20$.}
\label{fg:cxn_20}
\end{center}
\end{figure}
Choosing the dynamical scales $\mu_R=\mu_F = Q_i = \sqrt{-q_i^2}$ at
each proton leg, the QCD corrections are of moderate size, i.e.~at the
10--15\% level as can be inferred from Fig.~\ref{fg:k_vbf}\footnote{Fig.~\ref{fg:k_vbf} includes the corresponding LO and NLO scale-uncertainty bands, too. The K factors for $\tan\beta = 2,20$ are hardly distinguishable.}. The relative
QCD corrections develop only a minor dependence on the value of
$\tan\beta$. The related scale uncertainties determined by the 7-point
method are shown in Figs.~\ref{fg:sc_vbf1}, \ref{fg:sc_vbf2}. The NLO
error band lies within the LO one and reduces to residual uncertainties
to a level of less than 2--10\%. This uncertainty needs to be added to
the one of the SFA itself. The cross section of the VBF
does not develop a relevant dependence on the value of $\tan\beta$,
since it is dominated by the gauge-coupling contributions.

\section{Results} \label{sc:results}
Using the above computations we now present and compare the different charged-Higgs pair production processes and their corresponding uncertainties. 
We have adopted the PDF4LHC15 NLO PDFs \cite{pdf4lhc15} for the determination of
the PDF+$\alpha_s$ uncertainties combined with the scale dependence
discussed before. 
We have added the scale and PDF+$\alpha_s$ uncertainties {\it linearly}, since the scale dependence does not  show a significant dependence on the chosen error PDF.
The results are presented
in Figs.~\ref{fg:cxn_2} and \ref{fg:cxn_20} for the two scenarios with
$\tan\beta=2$ and 20, respectively. The Drell--Yan-like Higgs-pair
production mode dominates by one order of magnitude over the VBF and
gluon-fusion processes\footnote{Charged Di-Higgs-strahlung $q\bar q\to
H^+H^- + Z/W$ is expected to be further suppressed.}. The cross section
of the Drell--Yan-like process reaches a couple of $fb$ for smaller
charged Higgs masses and drops below the $ab$ range for charged Higgs
masses above 1.5 TeV for both scenarios, while the VBF and gluon-fusion
channels stay in the sub-$fb$ region with the VBF being the second
largest production process in a significant part of the mass range.

The PDF+$\alpha_s$ uncertainties of the Drell--Yan-like production mode
amount to about 2--17\%, thus combining with the scale uncertainties to a
total theoretical uncertainty of about 4--22\% depending on the charged
Higgs mass. The VBF develops PDF+$\alpha_s$ uncertainties of about 2--7\%
and combines with the scale dependence to about 3--15\% total
theoretical uncertainties. These are, however, modified by the
uncertainties induced by the SFA. The gluon-fusion process is dominantly
affected by its scale uncertainties of 20--25\%, while the
PDF+$\alpha_s$ uncertainties amount to about 4--20\% thus yielding 25--45\%
theoretical uncertainties in total. However, these are modified when
adding the missing finite top-mass effects of about 20--30\% at NLO. The
gluon-fusion cross-section is suppressed for large values of $\tan\beta$
in the 2HDM of type I.

\section{Conclusions} \label{sc:conclusions}
In this work we have analyzed charged Higgs-boson pair production at the
LHC. These production modes of novel charged Higgs states in extended Higgs
sectors exhibit direct access to some trilinear Higgs couplings and thus
to parts of the BSM Higgs potential. This analysis has been performed
within the 2HDM of type I.

The results for the DY-like and VBF charged-Higgs pair production are dominated by the charged-Higgs gauge couplings, while the contribution of the charged-Higgs self interactions are subleading in general. Therefore, our results are quite robust in the parameter space of the 2HDM. The results of the gluon fusion process dominantly depend on the top- and bottom-quark Yukawa couplings. Note that the NLO corrections added in this work are only valid for scenarios where the top-Yukawa coupling is dominant.

Although the cross sections for these
processes are small, i.e.~in the few-$fb$ down to the below-$ab$ range,
it will be relevant to search for these charged-Higgs production modes.
This work improved all contributing processes to the NLO QCD level thus
allowing for a refined analysis of the associated theoretical
uncertainties. The latter were determined by combining the scale
uncertainties by means of the 7-point method related to the
factorization and renormalization scale choices and the PDF+$\alpha_s$
uncertainties emerging from the PDFs. For the final combination of the
uncertainties, we adopted the NLO PDF4LHC15 PDFs and their associated
error PDF sets \cite{pdf4lhc15}, since the new PDF4LHC21 sets do not
provide LO and NLO PDFs \cite{pdf4lhc21}.

The dominant charged Higgs pair production mode is the Drell--Yan like
process with photon and $Z$-boson exchange in the $s$-channel. This
process is independent of the 2HDM scenario, since only gauge couplings
contribute (The bottom-Yukawa induced process $b\bar b\to H^+H^-$ is
suppressed by several orders of magnitude.). This results in residual
total theoretical uncertainties of about 4--22\% for the leading
Drell--Yan-like mode, while the uncertainties of the subleading VBF and
gluon-fusion amount to about 3--15\% and 25--45\%,
respectively\footnote{The upper ends of the uncertainty bands emerge for
large charged Higgs masses, where the PDFs contribute at large
Bjorken-$x$ values.}. The gluon-fusion cross section is strongly
suppressed for large values of $\tan\beta$ within the 2HDM of type I,
thus making VBF the second largest charged-Higgs production mode. \\

\noindent
{\bf Acknowledgments} \\
The authors are indebted to T.~Biek\"{o}tter and K.~Elyaouti for helpful correspondence and to E.~Vryonidou for useful discussions about Ref.~\cite{Hespel:2014sla}. We are grateful to S.~Moretti for sharing his LO code for vector-boson fusion used in Ref.~\cite{moretti} at very early stages of this work that allowed to pin down the sources of discrepancies. The research of L.B.~and Y.Y.~is supported by the Swiss National Science Foundation (SNSF).


\end{document}